\documentclass[aip,
graphicx,
reprint,
amsmath,
amssymb
]{revtex4-1}
\newcommand{\angstrom}{\text{\normalfont\AA}}
\providecommand{\abs}[1]{\lvert#1\rvert}
\draft 

\usepackage{bm}
\usepackage{graphicx,epstopdf}
\usepackage{color}

\begin{document}

\title{Simulating rare events using a Weighted Ensemble-based string method } 



\author{Joshua L. Adelman}
\email{jla65@pitt.edu}
\affiliation{Department of Biological Sciences, University of Pittsburgh, Pittsburgh, Pennsylvania 15260, USA}
\author{Michael Grabe}
\email{mdgrabe@pitt.edu}
\affiliation{Department of Biological Sciences, University of Pittsburgh, Pittsburgh, Pennsylvania 15260, USA}
\affiliation{Department of Computational \& Systems Biology, University of Pittsburgh, Pittsburgh, Pennsylvania 15260, USA}


\date{\today}

\begin{abstract}
We introduce an extension to the Weighted Ensemble (WE) path sampling method to restrict sampling to a one dimensional path through a high dimensional phase space.
Our method, which is based on the finite-temperature string method, permits efficient sampling of both equilibrium and non-equilibrium systems.
Sampling obtained from the WE method guides the adaptive refinement of a Voronoi tessellation of order parameter space, whose generating points, upon convergence, coincide with the principle reaction pathway.   
We demonstrate the application of this method to several simple, two-dimensional models of driven Brownian motion and to the conformational change of the nitrogen regulatory protein C receiver domain using an elastic network model.
The simplicity of the two-dimensional models allows us to directly compare the efficiency of the WE method to conventional brute force simulations and other path sampling algorithms, while the example of protein conformational change demonstrates how the method can be used to efficiently study transitions in the space of many collective variables.
\end{abstract}


\maketitle 

\section{Introduction}

Molecular simulations can provide deep insight into the mechanisms of physical processes, in part, because they inherently possess a spatial and temporal resolution that is unmatched by most experimental techniques.  
Unfortunately, many physical and biological processes such as chemical reactions, nucleation, protein conformational changes, and ligand binding occur on timescales that are inaccessible to conventional brute-force simulation methods.
Due to this shortcoming, there has been broad interest in developing methods that augment conventional simulations to allow them to capture rare events in a reasonable time frame\cite{Dellago1998Transition-path,Woolf1998Path-corrected-,Erp2003A-novel-path-sa,Faradjian2004Computing-time-,Allen2005Sampling-rare-switch,Warmflash2007Umbrella-sampli}, several of which are reviewed in Ref.~\onlinecite{Bolhuis2002Transition-path,Dellago2009Transition-path,Zwier2010Reaching-biolog}.
Almost all of these methods rely on enhancing sampling in a reduced set of collective coordinates that span the important regions of the high dimensional phase space.
The computational effort is thereby focused on sampling transition regions, which would otherwise be visited infrequently, if at all, in conventional simulations. 
The success of rare event sampling often hinges on the particular progress coordinate or order parameter used to discriminate movement along the transition, and choosing the proper progress coordinate is a non-trivial task.  
In fact, it is likely that most processes can only be described by multiple progress coordinates, which further complicates identification of the appropriate pathway through phase space.
The need to incorporate additional progress coordinates also drastically increases the computational demand since the cost scales like the power of the number of progress coordinates. 

Even if a transition occurs via a convoluted set of steps requiring multiple progress coordinates to describe the process, it is generally thought that the reaction can be meaningfully represented by a chain of connected nodes or beads (referred to as images) that define a path\cite{Elber1987A-method-for-determi,Fischer1992Conjugate-peak-,Henkelman2000Improved-tangen,Weinan2002String-method-f}. 
Methods built around this idea provide an elegant solution to the increased phase space needed to explore multiple directions because they are able to track an arbitrary number of progress coordinates while restraining the sampling to effectively one dimension. 
One such realization of this approach is the string method\cite{Weinan2002String-method-f}.
Since its advent, a number of variants of the string method have been developed to evaluate the transition pathways of complex systems\cite{Maragliano2006String-method-i,Weinan2005Finite-temperature-s}.
Additionally, the basic framework of the string method has been integrated with a variety of other sampling procedures\cite{Pan2008Finding-transition-p,Pan2008Building-Markov,Vanden-Eijnden2009Revisiting-the-finit,Dickson2009Nonequilibrium-umbre,Rogal2010The-reweighted-path-,Johnson2012Characterization-of-,Dickson2012Unrestrained-Co}.
While string methods have been applied to a diverse set of problems\cite{Ren2005Transition-pathways-,Miller2007Solvent-coarse-,Ovchinnikov2011A-conformational-tra}, reactions that occur via many, conformationally distinct pathways\cite{Bhatt2010Heterogeneous-p,Adelman2011Simulations-of-the-a} may be poorly suited for string methods. 
Studies have shown, however, as we do here, that when the transition is confined to a few reaction channels, which can be explicitly accounted for, string methods still perform well.

Here, we merge a rare event sampling method known as the Weighted Ensemble (WE) method\cite{Huber1996Weighted-ensemble-Br} with a string method, and we show that this combined approach performs both accurately and efficiently.
WE sampling is a rigorous method for sampling both equilibrium and non-equilibrium systems even in the presence of long-lived intermediate states\cite{Zhang2007Efficient-and-verifi,Zhang2010The-weighted-ensembl,Bhatt2010Steady-state-simulat}.
From a practical perspective, WE sampling is easy to implement and straightforward to parallelize across large computational resources.
The WE method has been previously used with Brownian dynamics simulations to study protein binding\cite{Huber1996Weighted-ensemble-Br,Rojnuckarin2000Bimolecular-rea} and protein folding\cite{Rojnuckarin1998Brownian-dynami}, and it has been used with explicit solvent molecular dynamics to investigate molecular association events\cite{Zwier2011Efficient-Expli}.
Our group, and others, have also extended the method to examine large-scale conformational transitions in biomolecular systems using coarse-grained models\cite{Zhang2007Efficient-and-verifi,Bhatt2010Heterogeneous-p,Adelman2011Simulations-of-the-a}.
Augmenting WE sampling with a string method yields not only the transition pathway represented by the converged string, but also a dynamically exact representation of the path ensemble along the string, from which steady-state distributions and kinetic information can be extracted from a single simulation.

We illustrate the WE-based string method by applying it to several example systems of varying degrees of complexity.
First, we examine two different nonequilibrium steady-state processes.
The first of these is a Brownian particle in a unidirectional flow on a periodic two-dimensional surface\cite{Dickson2009Nonequilibrium-umbre}.
Second, we examine a driven Brownian particle on a two-dimensional potential surface with intermediate metastable states along distinct forward and backward pathways\cite{Dickson2009Separating-forward-a}.
Finally, we apply the method to the equilibrium conformational transition of the nitrogen regulatory protein C receiver domain using a two-state elastic network model\cite{Pan2008Finding-transition-p,Vanden-Eijnden2009Revisiting-the-finit}.
All three systems have been previously studied with other string methods, facilitating a direct comparison with the WE-based algorithm.
Moreover, the low dimensionality of the first two examples makes it possible to rigorously sample each system with conventional simulation to test the accuracy of our method and serve as an additional benchmark for comparing specifically against nonequilibrium umbrella sampling (NEUS)\cite{Warmflash2007Umbrella-sampli,Dickson2009Nonequilibrium-umbre,Dickson2009Separating-forward-a}. 
Both WE and NEUS methods are able to calculate steady-state rates and distributions, even in the presence of long-lived intermediates, and the clarity of the framework developed in Refs.~\onlinecite{Dickson2009Nonequilibrium-umbre} and \onlinecite{Dickson2009Separating-forward-a} for comparing NEUS to conventional sampling has enabled us to include WE in this comparison. 
In all cases, the WE-based string method obtains high-fidelity estimates of steady-state distributions and rates using comparable, or less, computational resources than other string based methods. 
Thus, the procedure that we present here lays the groundwork for applying WE sampling to a broad range of problems in which many progress coordinates are required to efficiently partition phase space along a transition pathway.

\section{Methods}

\subsection{Weighted Ensemble path sampling}
Weighted Ensemble sampling is a general and rigorous method for simulating rare events\cite{Huber1996Weighted-ensemble-Br,Zhang2007Efficient-and-verifi}.   
This is accomplished via a resampling protocol that partitions the progress coordinate space of the system into non-overlapping bins that can be defined in an arbitrary number of dimensions\cite{Zhang2010The-weighted-ensembl}.
Here, resampling refers to a scheme that generates an alternative, but equivalent, statistical sample of a system's phase space, which in the context of WE sampling is accomplished as follows.

Multiple replicas of the system are initiated from some initial distribution of conformations.
Each replica is assigned a weight, such that the sum of the weights of all of the replicas in the system is unity.
In the simplest case this initial distribution of replicas may consist of $N_{rep}$ replicas of equal weight with identical coordinates, but randomly assigned velocities.
Each replica is then simulated independently for a short time interval $\tau$ during which it is allowed to explore conformational space following its natural dynamics.
At the end of this interval, every replica is assigned to a bin based upon its instantaneous coordinates at the end of the time interval. 

While assigning the configurations of the system into bins could be done using the full configurational space of the system, in practice the progress coordinates represent a set of collective coordinates in a reduced sub-space. 
Let $\bm{x}$ specify the configuration of the full system, and let $\bm{\theta}(\bm{x})$ be the mapping of $\bm{x}$ into the progress coordinate space.
The progress coordinates must be chosen to discriminate the product and reactant states, but need not specify a true reaction coordinate.

Within each bin, the number of replicas is held constant, or nearly constant, by occasionally replicating or terminating the replicas in that bin. 
If an occupied bin contains fewer than the target number of replicas, one or more of the replicas are selected via a statistical procedure and are split into $M$ new copies that each carry a fraction $1/M$ of the probability of the parent.
Conversely, if a bin contains more than the target number of replicas, a culling procedure removes excess replicas.
The weight of the culled replicas is redistributed to a subset of the remaining copies in that bin.  
These changes to the number of replicas within a bin are carried out in a way that does not bias the underlying dynamics\cite{Huber1996Weighted-ensemble-Br,Zhang2010The-weighted-ensembl}.
Metastable regions that would typically accumulate large numbers of replicas and consume a large fraction of the computational effort required to simulate the system are not over-populated because of the culling process.
Conversely, the conformational space atop a barrier between states that would normally be poorly sampled using conventional simulations is enriched with many low-weight replicas. 
Details of the replication and termination protocol can be found elsewhere\cite{Huber1996Weighted-ensemble-Br}. 

\subsection{String Method}

As in previous WE simulations, phase space is divided into non-overlapping regions or bins defined by a set of progress coordinates.
However, instead of a partition with the number of dimensions of the progress coordinate space, the bins are constructed as a string of Voronoi cells in a single dimension (the arclength along the string), embedded in the higher dimensionality progress coordinate space.
The string then evolves in time based on the dynamics of the replicas to follow the principle pathway through the conformational space.
Our implementation of a string method, adapted for WE sampling, closely follows the algorithm of the finite-temperature string (FTS) method suggested by Vanden-Eijnden and Venturoli\cite{Vanden-Eijnden2009Revisiting-the-finit}.

First, an initial path is constructed between two regions of phase space using a set of progress coordinates that are assumed to be sufficient to describe the transition between regions.  
A discretized set of evenly spaced images along the path, $\bm{\varphi}_{\alpha}$, partition the progress coordinate space into bins, $B_{\alpha}$.
The subscript $\alpha=1,... ,N_{\text{im}}$, where $N_{\text{im}}$ is the total number of images along the string. 
The string can be thought of as a spline with the images being the nodes connecting the individual segments.  
In practice, we use a linear interpolation to initialize the string with evenly spaced images. 
Each replica specified by $\bm{x}$ is mapped into progress coordinate space via the transformation $\bm{\theta}(\bm{x})$ and then assigned to the closest bin $B_{\alpha}$ by calculating the distance to all images $\bm{\varphi}_{\alpha}$. 
The image $\bm{\varphi}_{\alpha}$ is therefore the generator of the Voronoi cell corresponding to the bin $B_{\alpha}$.
Distance measurements in progress coordinate space require the use of the appropriate metric\cite{Vanden-Eijnden2009Revisiting-the-finit}; 
however, once the metric has been defined, assigning a replica to a bin is straightforward, regardless of the dimensionality of that space.
A detailed discussion of how to estimate the metric for a set of collective coordinates is given in Ref.~\onlinecite{Vanden-Eijnden2009Revisiting-the-finit}.
The boundaries separating bins, which can be quite complicated in high-dimensional spaces, need not be computed. 

The string images are then adaptively updated according to the following algorithm:

\begin{enumerate}
\item Simulate the system to explore phase space for a time $T_{\text{move}}$, which is generally an integral number of $\tau$. 
	During this period, replicas are free to move between bins, unlike in the finite-temperature string method\cite{Vanden-Eijnden2009Revisiting-the-finit} or 
	nonequilibrium umbrella sampling\cite{Warmflash2007Umbrella-sampli,Dickson2009Nonequilibrium-umbre}.

\item Compute the average position of all replicas in each bin over the time $T_{\text{avg}}$ according to:
	\begin{eqnarray}
	\left<\bm{\theta}_{\alpha}(\bm{x})\right> = \frac
		{  \sum_{i=1}^{N_r}      w_{i}\bm{\theta}(\bm{x}_{i}) h_{\alpha}(\bm{\theta}(\bm{x}_{i}))}
		{  \sum_{i=1}^{N_r}      w_{i} h_{\alpha}(\bm{\theta}(\bm{x}_{i})) },
		\label{eq:average}
	\end{eqnarray}
	where $i$ indicates the $i^{th}$ of $N_r$ replicas simulated in this interval, $w_{i}$ is the weight of replica $i$
	and $h_{\alpha}(\bm{\theta}(\bm{x}_{i}))$ is an indicator function for Voronoi cell $\alpha$, which assumes a value of $1$ if replica $i$ is in bin $\alpha$, and $0$ otherwise.
	
\item Update the images along the strings, moving them toward the average positions within each bin according to: 
	\begin{eqnarray}
	\bm{{\varphi}}_{\alpha}^{*} = \bm{{\varphi}}_{\alpha}^{n} - \zeta \left( \bm{{\varphi}}_{\alpha}^{n} - \left<\bm{\theta}_{\alpha}\right> \right) + \bm{r}_{\alpha}^{*},
	\label{eq:string_update}
	\end{eqnarray}
	where $\bm{{\varphi}}_{\alpha}^{*}$ are the updated images, $\bm{{\varphi}}_{\alpha}^{n}$ are the previous position of the images, 
	$\zeta>0$ controls how rapidly the image moves toward the current estimate of the average position in the bin, and $\bm{r}_{\alpha}^{*}$ is a smoothing term discussed below.
	 
\item Redistribute images uniformly along the arc length of the string by fitting a piece-wise linear spline through all $\bm{{\varphi}}_{\alpha}^{*}$ and then spacing images equidistantly along the curve. 
This reparameterization of the string ensures for a given update iteration, $n$,
\begin{eqnarray}
\left| \bm{{\varphi}}_{\alpha + 1}^{n} - \bm{{\varphi}}_{\alpha}^{n} \right| = \left| \bm{{\varphi}}_{\alpha}^{n} - \bm{{\varphi}}_{\alpha - 1}^{n} \right|.
\end{eqnarray}
This final step prevents the images from drifting toward nearby metastable states, which would compromise the efficient sampling of the transition regions.
Relabel the shifted  $\bm{{\varphi}}_{\alpha}^{*}$ as $\bm{{\varphi}}_{\alpha}^{n+1}$. 
\end{enumerate}

Steps 1-4 are iterated until the images defining the string are stationary. 

While this rough outline defines the method, there are additional details employed in the current study.
For instance, system configurations, $\bm{x}_{i}$, used to identify bin centers in Eq. \ref{eq:average} are taken as the coordinates at the end of each WE simulation of length $\tau$. 
While this choice of averaging works well for the systems presented here, more or less frequent configuration snapshots should be averaged depending on the length of $\tau$ compared to the natural decorrelation time of the system within a bin.
The key is to ensure statistical independence of configurations, while not averaging too infrequently.
Additionally, using only the last $T_{\text{avg}}/\tau$ iterations in calculating the average position of replicas within a bin limits the effect of early sampling near the initial position of the string.
This is desirable since the weight of replicas within each bin can change rapidly early in the simulation before reaching steady state, thus biasing the calculation of the average.

Smoothing the string as it evolves is essential to prevent kinks and other pathological shapes from developing\cite{Ren2005Transition-pathways-}. 
We accomplish smoothing through manipulating the term $\bm{r}_{\alpha}^{*}$ in Eq. \ref{eq:string_update} using one of two different methods. 
First, for the protein conformational change example (Sec.~\ref{sec:ntrc}), we define
\begin{eqnarray}
\bm{r}_{\alpha}^{*} = \kappa^n \left( \bm{{\varphi}}_{\alpha + 1}^{*} + \bm{{\varphi}}_{\alpha - 1}^{*} - 2\bm{{\varphi}}_{\alpha}^{*}\right) , 
\label{eq:implicit_smooth}
\end{eqnarray}
for $\alpha = 1{\ldots}N-1$, $\kappa^n = {\kappa}N_{im}\zeta$ and $\bm{r}_{0}^{*} = \bm{r}_{N}^{*}=0$.
This term adds an effective elasticity to the string that prevents the image from moving to $\left<\bm{\theta}_{\alpha}\right>$ if it causes large kink angles between the current image and its neighbors. 
The parameter $\kappa$, which is positive definite, controls how aggressively the string is smoothed\cite{Vanden-Eijnden2009Revisiting-the-finit}. 
Alternatively the updated positions can be smoothed using a multidimensional curve fitting procedure\cite{Zhu2010Pore-opening-and-clo}.
This procedure was used for the examples in both Sec.~\ref{sec:dpp} and \ref{sec:drp}.
First, $\bm{{\varphi}}_{\alpha}^{*}$ are calculated with $\bm{r}_{\alpha}^{*} = 0$ for all $\alpha$.
Then a smooth continuous path connecting $\bm{{\varphi}}_{0}^{*}$ and $\bm{{\varphi}}_{N}^{*}$ is generated by fitting 
\begin{eqnarray}
\label{eq:multidim_string}
\bm{{\varphi}}_{\alpha}^{*,\text{cur}}(\lambda) = \bm{{\varphi}}_{0}^{*} + \left(\bm{{\varphi}}_{N}^{*} - \bm{{\varphi}}_{0}^{*} \right)\lambda+ \nonumber \\
\sum_{i=1}^{N_{\text{dim}}} \sum_{j=1}^{P} \sigma_{i,j}\sin(j{\pi}\lambda) \cdot  \hat{e}_i , 
\end{eqnarray}
to $\bm{{\varphi}}_{\alpha}^{*}$ by varying $\lambda$ over the range $[0,1]$. 
Here, $N_{\text{dim}}$ is the dimensionality of the progress coordinate space, $\hat{e}_i$ is the unit vector of the $i^{th}$ progress coordinate and $\sigma_{i,j}$ are the coefficients of $P$ sinusoidal basis functions in each dimension.
The superscript $cur$ indicates that $\bm{{\varphi}}_{\alpha}^{*,\text{cur}}$ on the left hand side is calculated by the curve fitting procedure.
The parameters $\sigma_{i,j}$ and ${\lambda}_{\alpha}$ are selected to minimize
\begin{eqnarray}
{\chi}^2 = \sum_{\alpha=0}^{N_{im}} \left| \bm{{\varphi}}_{\alpha}^{*,\text{cur}}({\lambda}_{\alpha}) - \bm{{\varphi}}_{\alpha}^{*} \right|^2.
\end{eqnarray}
Details of the optimization procedure to minimize ${\chi}^2$ are given in the supplementary materials of Ref.~\onlinecite{Zhu2010Pore-opening-and-clo}.

\subsection{\label{sec:reweighting}Re-weighting Scheme}

In the original formulation of the WE method, the initial distribution of weights gradually redistributes throughout the sampled conformational space and reaches either an equilibrium or non-equilibrium steady state after some finite number of iterations\cite{Huber1996Weighted-ensemble-Br}.
In the presence of metastable intermediates, the time for the system to relax to the correct steady-state distribution of weights can be very long, which hampers the efficiency of the method 
\cite{Bhatt2010Steady-state-simulat}.
It was shown previously that this relaxation time can be dramatically reduced through a re-weighting procedure that used estimates of the rate of transitions between bins to solve for the steady-state distribution \cite{Bhatt2010Steady-state-simulat}.
Briefly, since the dynamics of the individual replicas in a WE simulation are unbiased, the transition rates between the bins can be used to estimate the elements of a transition matrix $\bm{k}$.
A complete specification of the transition rates allows one to solve for the steady-state probabilities using
\begin{eqnarray}
\label{eq:reweight}
\frac{dP_i}{dt}=\sum_j k_{j,i}P_j - \sum_j k_{i,j}P_i = 0,
\end{eqnarray}
where $P_i$ is the probability associated with bin $i$, and $k_{i,j}$ is the rate of transitions from bin $i$ to $j$ and $\sum_i{P_i}=1$.
The new estimate of $P_i$ is then used to re-weight the replicas in bin $i$, where the weight is distributed among the replicas in the bin in proportion to their weights before re-weighting.
For the example in Sec.~\ref{sec:ntrc}, which involves equilibrium sampling, we use a different re-weighting procedure that uses detailed balance to constrain the estimate of the steady state distribution\cite{Lettieri2012Simultaneous-computa}.

Unlike previous WE studies\cite{Bhatt2010Steady-state-simulat,Adelman2011Simulations-of-the-a}, which employed a single re-weighting step based on a short period of initial sampling, we carry out multiple re-weighting steps at periodic intervals $T_{\text{rw}}$ as suggested in Refs.~\onlinecite{Dickson2009Separating-forward-a,Vanden-Eijnden2009Exact-rate-calculati,Dickson2011Flow-dependent-,Lettieri2012Simultaneous-computa}
The rates are calculated over a window spanning $T_{\text{ravg}}$ simulation steps.
In practice we use a fractional window that extends from the current simulation time back $T_{\text{ravg}}/\tau$ iterations of the WE resampling procedure, which discards inaccurate contributions to the rates early in the simulation.
That fraction is fixed during the re-weighting phase to $T_{\text{ravg}}/{{N_\tau}\tau}$, where $N_\tau$ is the current iteration number. 

When the steady state re-weighting method is used in a simulation, we separate our simulation protocol into distinct phases. 
We apply the re-weighting protcol only during the initial phase, early in the simulation, since it efficiently relaxes the system away from the initial inaccurate distribution of weights toward the correct steady state distribution.
We find, however, that the standard WE method obtains a higher overall level of accuracy as steady state is approached.
The reason for this increased accuracy is related to statistical net zero flux between bins emerging naturally at steady state with standard WE, rather than being enforced based on potentially inaccurate rates determined by the re-weighting scheme.
Similar observations were made by Dickson, \emph{et al.} for the NEUS method, and they also only use a global re-weighting scheme early in the simulations \cite{Dickson2009Separating-forward-a}.

\subsection{\label{sec:calc_rates}Calculating rates}

It is of interest to use simulation to calculate the rates for a system to interconvert between distinct states $A$ and $B$.
This can be achieved from very long conventional simulations by observing many transitions between $A$ and $B$ and separating the $A \rightarrow B$ transitions from the $B \rightarrow A$ transitions.
Explicitly, at some time $t$, a trajectory that had last been in state $A$ is assigned to state $A$ ($\mathcal{S}_A$) at time $t$; if it had last been in state $B$, it is labeled as belonging to state $B$ ($\mathcal{S}_B$).
Given this partition, the reaction rate from state $A \rightarrow B$ or from $B \rightarrow A$ is given by \cite{Vanden-Eijnden2009Exact-rate-calculati}: 
\begin{eqnarray}
k_{A,B} = \lim_{T\to \infty} \frac{N^{T}_{A,B}}{T_{A}} , k_{B,A} = \lim_{T\to \infty} \frac{N^{T}_{B,A}}{T_{B}},
\label{eq:rate}
\end{eqnarray}
where $T_A$ and $T_B$ are the total time that a trajectory has been assigned to state $A$ or $B$, respectively, during the interval $[0,T]$.
During the same interval, $N^{T}_{A,B}$ is the number of times a trajectory assigned to $A$ switched to being assigned to $B$, and $N^{T}_{B,A}$ enumerates the reverse transition.

An alternative but equivalent definition of the reaction rate based on fluxes into $A$ and $B$ is given by\cite{Allen2005Sampling-rare-switch}
\begin{eqnarray}
k_{A,B} = \frac{\overline{\Phi}_{B|\mathcal{S}_A}}{\overline{h}_A}, k_{B,A} = \frac{\overline{\Phi}_{A|\mathcal{S}_B}}{\overline{h}_B} ,
\label{eq:flux_rate}
\end{eqnarray}
where $\Phi_{B|\mathcal{S}_A}$ ($\Phi_{A|\mathcal{S}_B}$) is the flux into state $B$ ($A$) from trajectories that originated in state $A$ ($B$), and $h_A$ ($h_B$) is a history dependent indicator function that is equal to 1 if the trajectory was more recently in state $A$ ($B$) than in $B$ ($A$), and zero otherwise.
The indicator function serves to assign trajectories to either state $A$ or $B$ as above.
Overbars indicate a time averaged quantity.

The original formulation of the Weighted Ensemble method, in which replicas originating from a reactant state (state $A$) were reintroduced into that state upon crossing the product surface (state $B$), effectively isolated members of the trajectory assigned to a single state  \cite{Huber1996Weighted-ensemble-Br,Bhatt2010Steady-state-simulat}.
In principle, WE does not require one to run simulations with this feedback scheme, since it properly accounts for a dynamical trajectory's history\cite{Lettieri2012Simultaneous-computa}.
Thus, a state can be unambiguously assigned (assuming that all trajectories are initiated from $A$ or $B$ -- otherwise there will be some transient period when a trajectory is unassigned) to every replica enabling one to calculate the requisite  fluxes in Eq.~\ref{eq:flux_rate}.

Dickson and colleagues introduced a dual-direction scheme that uses an extended bin space to split the forward and backward path ensembles\cite{Dickson2009Separating-forward-a}.
In the standard nomenclature of a WE simulation with $N_{\text{dim}}$ progress coordinates, one adds an additional progress coordinate which labels each replica as either being assigned to the product or reactant state.
For the Voronoi-based division of the progress coordinate space used in this study, this additional progress coordinate modifies the distance metric used to assign a set of coordinates to a bin:
\begin{eqnarray}
\|\bm{x}-\bm{{\varphi}}_{\alpha}^{n}\| = 
\begin{cases}
\infty & \text{if } x_{m+1} \neq {\varphi}_{\alpha,m+1}^{n} \\
\sqrt{\sum\limits_{i=1}^{m} \left(x_i - {\varphi}_{\alpha,i}^{n}\right)^{2}} & \text{otherwise.}
\end{cases}
\label{eq:ext_distmetric}
\end{eqnarray}
A replica who had last visited $A$ is then infinitely far from the centers of the string associated with state $B$.
During a WE simulation then, $\overline{\Phi}_{B|\mathcal{S}_A}$ is calculated as the flux of probability from all Voronoi cells in $\mathcal{S}_A$ into the region defining state $B$.

\subsection{\label{sec:implementation} Implementation}

We have implemented all of the systems in Sec.~\ref{sec:examples} using open-source software written in Python.
The string method was implemented as a plugin for the Weighted Ensemble Simulation Toolkit with Parallelization and Analysis (WESTPA)\cite{west-pa}, which provides a general framework for performing and analyzing WE simulations.
The complete set of tools required to simulate the example systems, analyze the results and generate the figures found in this study are available at \url{https://simtk.org/home/westring}.
\section{\label{sec:examples}Examples}
\subsection{\label{sec:dpp}Periodic two-dimension system}
As an initial test of the string-based Weighted Ensemble sampling method, we apply it to a periodic two-dimensional system \cite{Dickson2009Nonequilibrium-umbre} with a potential surface defined by
\begin{eqnarray}
V(x,y) = \gamma\left[x - \frac{1}{2}\sin(2{\pi}y)\right]^{2} + \alpha \cos(2{\pi}y),
\label{eq:DicksonPeriodicPotential}
\end{eqnarray}
where $x$ and $y$ are the spatial coordinates and $\alpha$ and $\gamma$ are parameters that determine the shape of the potential.
The form of the potential creates a reaction pathway that depends nontrivially on both coordinates, $x$ and $y$.

Periodic boundary conditions are applied in the $y$ direction such that $y \in [0,1)$.
A constant external force ($\bm{F}_{ext}=F\hat{y}$) is applied along the $y$ direction, which drives the system out-of-equilibrium and generates a constant flux across the periodic boundary.
A particle on the potential evolves according to the over damped equation of motion in the presence of a random stochastic fluctuation. 
The discretized form of the equation is
\begin{eqnarray}
\bm{X}(t + {\delta}t) = \bm{X}(t) - \frac{{\delta}t}{m\xi}(\bm{\nabla_{X}}V - \bm{F}_{ext}) + \delta\bm{X}^G,
\label{eq:EqnMotion}
\end{eqnarray}
where $\bm{X} = (x,y)$, ${\delta}t$ is the time step, $\delta\bm{X}^G$ is a random displacement with zero mean and variance $2D{\delta}t$.
The diffusion coefficient, $D=(m{\beta}{\xi})^{-1}$ is defined in terms of the mass of the particle, $m$, the inverse temperature, $\beta$ and the friction coefficient, $\xi$.

In this work, ${\delta}t=0.002$, $\xi = 1.5$, $F = 1.8$, $\beta = 4.0$, $m = 1.0$ and $\gamma = 2.25$.
Parameters related to the string parameterization and WE sampling are given in Table~\ref{tab:2dper_params}.
The value of $\alpha$ modulates the height of the barrier spanning the periodic boundary.
Using two different values of $\alpha$, we consider a case where transitions are common ($\alpha=1.125$) and another where the barrier is higher and  transitions are rare ($\alpha=2.25$).
These parameters were selected to match the values used in Ref.~\onlinecite{Dickson2009Nonequilibrium-umbre}, although they differ from the values originally reported\cite{Dickson2012Erratum:-Nonequ}.

\begin{figure}
\includegraphics{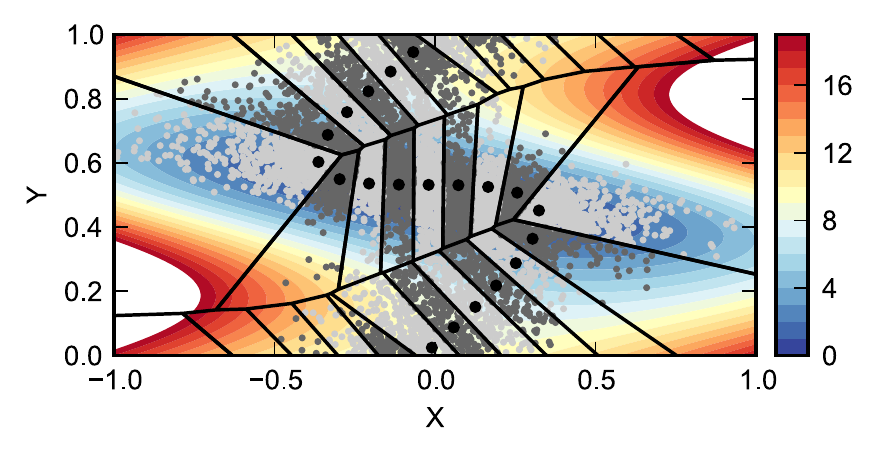}
\caption{
\label{fig:dpp_potstr}
The two-dimension periodic potential surface for ${\alpha}=1.125$, with the converged path of 20 centers (black dots) and corresponding Voronoi cells.
Contour lines are separated by $kT$, and the corresponding color scale is shown in the same units.
In each Voronoi cell, a random sample of the instantaneous positions visited by the replicas is shown in alternating light and dark gray dots for clarity. 
}
\end{figure}

Figure~\ref{fig:dpp_potstr} shows the converged string for the common transition case after 1000$\tau$.
The string was initialized as a vertical line between $(0,0.05)$ and $(0,0.95)$.
Updates to the string were performed using the multidimensional curve fitting procedure (Eq.~\ref{eq:multidim_string}) with $P=2$.
For both conventional and WE simulations, we generate a projection of the steady-state probability distribution onto the $y$ axis as the simulations progress in time.
These projections are then used to analyze the convergence properties of each simulation method compared to a well-converged target distribution.
The discrepancy between the target distribution and the simulated distribution accumulated to a particular time point on a logarithmic scale is      
\begin{eqnarray}
\text{error} = \left(  \frac{1}{n} \sum^{n}_{i=1}E^{2}_{i} \right)^{\frac{1}{2}},
\label{eq:dpp_error_a}
\end{eqnarray}
where
\begin{eqnarray}
E_i = 
\begin{cases}
\text{log } P(i) - \text{log } P_{t}(i) & P(i) \neq 0 \\
\text{log } 1/T - \text{log } P_{t}(i) & P(i) = 0 \\
\end{cases}
\label{eq:dpp_error_b}
\end{eqnarray}
Here, $P(i)$ is the normalized steady-state probability distribution projected onto the $y$ axis, $P_{t}(i)$ is the target distribution and $T$ is the total time of the complete simulation.
The histograms used to construct the distributions are obtained by dividing the $y$ axis between 0 and 1 into $n=100$ windows of equal width. 
The alternative definition of the error for $P(i) = 0$ is necessary to ensure that the error is still finite when bins have yet to accumulate any samples in them.
The impact of this choice is only significant for the conventional sampling case when ${\alpha}=2.25$ since these simulations require a non-negligible amount of time to reach the windows at the top of the higher barrier.
In both sets of WE simulations, all of the histogram windows are populated almost immediately, even if the initial estimate of the probability is inaccurate. 

\begin{figure}
\includegraphics{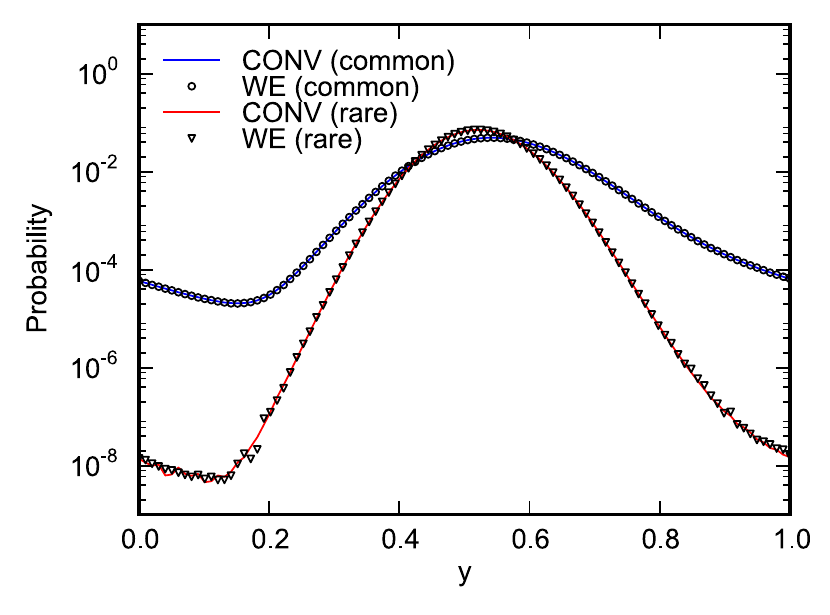}
\caption{
\label{fig:dpp_distributions}
Projections of the steady-state distributions for the two-dimensional periodic potential onto the $y$ axis.
Solid lines show the distributions for the common and rare parameter sets obtained using conventional sampling.}
\end{figure}

The final distributions show excellent agreement between the WE simulations and the long conventional target simulations for both values of $\alpha$ (Fig.~\ref{fig:dpp_distributions}).
At intermediate times, Fig.~\ref{fig:dpp_error} shows that for both choices of $\alpha$, the WE simulations converge to the correct distribution more rapidly than the respective conventional simulations.
For $\alpha=1.125$, the conventional simulations take approximately five times longer to converge to the same level of accuracy as the WE simulations.
For the case where $\alpha=2.25$, the barrier is high enough that the conventional simulations require a significant number of steps to accumulate sampling in every histogram window.
During this phase in which a particle in the conventional simulations has not visited every window, the error is dominated by the term in Eq.~\ref{eq:dpp_error_b} corresponding to $P(i)=0$, decreasing slowly until all of the bins have been visited.
The WE simulations, conversely, populate every bin almost immediately and the total error decreases much more rapidly. 
As such, the methods display quite different convergence characteristics for the high barrier case, with the WE simulations converge to a specific error in at least an order-of-magnitude fewer steps than conventional simulations, and significantly faster for certain error values. 

\subsection{\label{sec:drp}Two-dimensional system with two pathways}

As a second example, we consider a two-dimensional ring-shaped potential with two distinct pathways \cite{Dickson2009Separating-forward-a} connecting a pair of metastable states.
A transition from one state to the other requires passing through a metastable intermediate, which makes this system more difficult to sample than the model examined in \ref{sec:dpp}.
Many complex systems contain metastable intermediates along the transition path, so such a test becomes important in ensuring the procedure's broader applicability.
The potential surface for this model is defined as 
\begin{eqnarray}
V(r,\theta) = {\alpha}(r-\gamma)^2 + {\chi}_1\cos(2\theta) - {\chi}_2\cos(4\theta).
\end{eqnarray}

Here, $r=(x^2 + y^2)^{1/2}$ and $\theta$ is the angle in radians measured counterclockwise from the $x$ axis. 
Particles on this surface evolve according to Eq.~\ref{eq:EqnMotion} with the same definition of the noise term and diffusion constant.
A constant external force, $\bm{F}_\text{ext} = -F\hat{\theta}/r$, drives the system out of equilibrium in a clockwise direction \cite{Dickson2012Erratum:-Separa}.
The parameters governing the dynamics and shape of the potential surface were selected to match those used in Ref.~\onlinecite{Dickson2009Separating-forward-a}, where $\alpha = 3.0$, $\gamma = 3.0$, ${\chi}_1 = 2.25$, ${\chi}_2 = 4.5$, $\xi = 1.5$, $F = 7.2$, and ${\delta}t = 0.005$.
The inverse temperature is varied between $\beta=1.0$ and $3.0$.
On this surface, we define two states, $A$ and $B$, which encompass the area within circles of radius $R=1.0$ centered at $(-\gamma, 0)$ and $(\gamma, 0)$, respectively. 
The external force creates two distinct pathways for the forward ($A{\to}B$) and backward ($B{\to}A$) transitions (Fig.~\ref{fig:drp_potstr}).

Parameters associated with the WE protocol for each inverse temperature are given in Table~\ref{tab:2drp_params}, and each string is initialized as a horizontal line connecting the centers of $A$ and $B$.
A total of $N_{\text{rep}}$ replicas are initiated at each of the positions $(-\gamma, 0)$, $(\gamma, 0)$, $(0,-\gamma)$ and $(0, \gamma)$, and are assigned a weight of $(1-3\delta)/N_{\text{rep}}$ at $(-\gamma, 0)$, and  $\delta/N_{\text{rep}}$ otherwise, where $\delta=1\times10^{-12}$.

The presence of metastable intermediate states centered at $(\gamma, 0)$ and $(0,-\gamma)$ along each path requires the use of the re-weighting scheme described in Section~\ref{sec:reweighting} to efficiently converge to the correct non-equilibrium steady state distribution, as our choice of initial weights starts the system far from steady state.
Projections of the steady state distribution for values of $\beta{\le}2.5$ onto the angular coordinate $\theta$ are shown for both conventional (CONV) and weighted ensemble (WE) simulation in Fig.~\ref{fig:drp_distributions}.
There is excellent correspondence between the distributions obtained from the conventional simulations and the WE method across the entire temperature range in terms of both the density in the metastable states as well as the barrier regions.
 
For this model system, we analyzed the convergence behavior of the reaction rates instead of the steady state distribution, following Ref.~\onlinecite{Dickson2009Separating-forward-a}.
Separating the ensemble of replicas transiting in each direction onto two strings using the dual-direction scheme described in Section~\ref{sec:calc_rates}, allowed us to calculate the reaction rate using Eq.~\ref{eq:flux_rate}. 
We then analyzed the performance of the WE method in calculating the reaction rate between states $A$ and $B$.
The WE simulations are compared against conventional simulations in which the rate is calculated using Eq.~\ref{eq:rate}.   
The error in the calculated rate is measured as
\begin{eqnarray}
\label{eq:drp_error}
\text{error}(t) = | \log k(t) - \log k_t |,
\end{eqnarray}
where $k_t$ is the target rate constant and $k(t)$ is the rate as measured at a time $t$ after the start of the simulation.
Target rate constants for $\beta{\le}2.5$ were calculated by averaging the rates obtained from ten conventional simulations, each longer than 200 times the mean first passage time (MFPT).
For ${\beta}=3.0$, the target rate constant was obtained by fitting the rates from the higher temperature target simulations to an Arrhenius form and extrapolating.
Errors as a function of aggregate simulation time are plotted for ${\beta}=1.5$ and $2.5$ in Fig.~\ref{fig:drp_error} and are the root mean squared averages of errors obtained from ten simulations.
In the case of the conventional simulation, errors and target rates are calculated from independent simulations. 
The curves corresponding to the WE simulations are a moving average over the five previous iterations, and rates are based on a moving average with a window size of 200$\tau$.
 
The performance of the WE simulations over the entire temperature range is shown in Fig.~\ref{fig:drp_rate_ord_mag} by plotting $T_X/\text{MFPT}$ as a function of $\beta$.
The amount of simulated time, $T_X$, required to obtain an error equal to $X$ is calculated using Eq.~\ref{eq:drp_error}.
Measuring the error, $X$, on a logarithmic scale, $T_1$ corresponds to the amount of time required to obtain an order-of-magnitude estimate of the rates ($\text{error} \sim 10^1$), while $T_{0.5}$ is the time required to get an estimate that is within a factor of three of the target ($\text{error} \sim 10^{0.5}\sim3$).  
When $T_{1}/\text{MFPT}=1$, the amount of time required to obtain an order-of-magnitude estimate of the rate is approximately equal to the amount of time required by a conventional simulation.    

Figure~\ref{fig:drp_rate_ord_mag} indicates that the WE method is significantly more efficient than the corresponding conventional simulation at all temperatures except for the highest (${\beta}=1.0$), in which particles on the potential can easily cross the barriers between states.
For $\beta=1.0$, the time required to relax away from the initial distribution of weights (approximately a $\delta$ function of probability in state $A$) and to the correct steady state probability distribution  is of the same order of time as the MFPT.
The comparative efficiency of WE to conventional sampling increases with decreasing temperature (increasing barrier height); 
WE obtains an order-of-magnitude estimate of the rate in nearly four orders-of-magnitude less time than the MFPT for $\beta=3.0$.

\subsection{\label{sec:ntrc}Elastic Network Model of the NTRC$^{r}$ protein domain}

Finally, we consider the allosteric transition between the inactive and active conformations of the nitrogen regulatory protein C receiver domain (NtrC$^{r}$).
This system has been studied previously using both all-atom\cite{Damjanovic2009Self-guided-Langevin,Lei2009Segmented-transition} and coarse-grained simulation\cite{Pan2008Finding-transition-p,Vanden-Eijnden2009Revisiting-the-finit}.
For the purpose of testing the WE string method, we choose to follow the latter approach, and guided by those studies, construct a two-state elastic network model of the protein.

Elastic network models provide a reduced representation of the protein, where only the C$_{\alpha}$ atom of the backbone is explicitly included.
The interaction between these coarse-grained sites is governed by harmonic potentials that stabilize one or more reference conformations, often the experimentally determined native conformation.
While elastic network models lack the chemical fidelity of all-atom models, they have been shown in some instances to recapitulate important conformational fluctuations of less simplistic models\cite{Adelman2010The-mechanical-prope,Lyman2008Systematic-multiscal,Yang2008Close-correspondence}.

Following the model construction detailed in Refs.~\onlinecite{Pan2008Finding-transition-p,Vanden-Eijnden2009Revisiting-the-finit}, we build a two-state elastic network model of NtrC$^{r}$ to study the transition from the inactive to active states of the protein upon phosphorylation.
These states are generated from the 124 C$_{\alpha}$ positions in the NMR structures\cite{Kern1999Structure-of-a-trans} (PDB IDs 1DC7 and 1DC8, for state $A$ and $B$, respectively) and are shown in Fig.~\ref{fig:ntrc_struct}.  
The potential energy of the protein is specified by its instantaneous conformation, $\bm{x} \in R^{3M}$, where $M$ is the number of residues in the model and $\bm{x}_i$ denotes the position of the i${^{\text{th}}}$ residue.
Specifically, 
\begin{eqnarray}
U(\bm{x}) = -\frac{1}{{\beta}_m}\text{ln}\left(e^{-{\beta}_{m}U^{A}(\bm{x})} +  e^{-{\beta}_{m}U^{B}(\bm{x})}\right) + U^{R}(\bm{x}),
\label{eq:TwoStateElasticPot}
\end{eqnarray}
where $U^{A}(\bm{x})$ and $U^{B}(\bm{x})$, defined below, are the individual elastic network model energies for reference states $A$ and $B$, respectively.
These single-well potentials are combined by exponential averaging, where the parameter ${\beta}_m$ controls the barrier height that separates the two states and thus the rate of transitions.
\begin{eqnarray}
U^{A}(\bm{x}) = \frac{1}{2}\sum^{M}_{ij}k_{ij}D^{A}_{ij}({\Delta}x_{ij} - {\Delta}x^{A}_{ij})^{2},
\end{eqnarray}
with an analogous definition for $U^{B}(\bm{x})$.
The distance between residues $i$ and $j$, ${\Delta}x_{ij}=\abs{\bm{x}_i - \bm{x}_j}$.
Distances between residues calculated from the reference state structures $A$ and $B$ are defined as ${\Delta}x^A_{ij}=\abs{\bm{x}^A_i - \bm{x}^A_j}$ and ${\Delta}x^B_{ij}=\abs{\bm{x}^B_i - \bm{x}^B_j}$, respectively.
The contact matrices $D^{A}_{ij}$ and $D^{B}_{ij}$ determine which pairs of residues are connected via harmonic linkages, are given by 
\begin{eqnarray}
D^{A}_{ij} = 
\begin{cases}
1, & {\Delta}x^{A}_{ij} < d^{A} \\
0, & \text{otherwise}
\end{cases}
\end{eqnarray}
and similarly for $D^{B}_{ij}$, where $d^{A}$ is a cutoff distance. 
The force constants, $k_{ij}$ are modulated by the difference in pairwise residue-residue distances in $A$ and $B$ as 
\begin{eqnarray}
k_{ij} = \text{min} \left( \frac{{\varepsilon}_{k}} {({\Delta}x^{A}_{ij} - {\Delta}x^{B}_{ij})^2}, k_{\text{max}}\right). 
\end{eqnarray}
The final term in Eq.~\ref{eq:TwoStateElasticPot} provides a hard core repulsion that prevents steric overlap between residues, and is given by  
\begin{eqnarray}
U^{R}(\bm{x}) = {\varepsilon}\sum^{M}_{\substack{i,j=1,\\i{\neq}j}}\left(\frac{\sigma}{{\Delta}x_{ij}}\right)^{12}.
\end{eqnarray}
The parameterization of the potential is the same as in Ref.~\onlinecite{Vanden-Eijnden2009Revisiting-the-finit}: 
$d^{A,B}=11.5\;{\angstrom}$, 
${\varepsilon}_{k}=0.5\;\text{kcal mol}^{-1}$, 
$k_{\text{max}} = 0.2\;\text{kcal mol}^{-1}{\angstrom}^{-2}$, 
${\varepsilon} = 1.0\;\text{kcal mol}^{-1}$, 
$\sigma = 1.0\;\angstrom$, 
${\beta}_m = 0.02\;\text{kcal mol}^{-1}$, 
and the masses of all particles, $m=100\;\text{amu}$.
As in Ref.~\onlinecite{Vanden-Eijnden2009Revisiting-the-finit}, we used a modified value of ${\beta}_m$ that differs from the original model given in Ref.~\onlinecite{Pan2008Finding-transition-p}, where
${\beta}_m$ was set to $0.005$ $\text{kcal mol}^{-1}$.

We then introduce a set of collective coordinates ${\bm{\theta}}(\bm{x})=R\bm{x} +\bm{X}$, which we will use to assign each conformation to an image along the string.
The coordinate ${\bm{\theta}}$ is of the same dimensionality as our original system, but removes the degeneracies due to translation and rotation of the system. 
In the above definition, $R$ is the rotation matrix and $\bm{X}$ is the translation vector that when applied to $\bm{x}$ minimizes $\left| \bm{\theta}(\bm{x}) - \bm{x}_{\text{ref}} \right|$.
The distance metric in the collective coordinate space is the root mean square deviation (RMSD) of the optimally aligned conformation $\bm{\theta}(\bm{x})$ with some reference coordinates of the protein, $\bm{x}_{\text{ref}}$.
Here, the RMSD is calculated using the fast quaternion-based characteristic polynomial method\cite{Theobald2005Rapid-calculation-of}, which does not require the explicit calculation of $R$.
The two residues at both the C- and N-Termni of the protein are highly mobile and are excluded from the RMSD calculation. 

The string is initiated as the linear interpolation between $A$ and $B$ using 40 images ($N_{im}=40$).
Forty replicas ($N_{rep}=40$) were initiated in the bins corresponding to $A$ and $B$ with each replica given equal weight.
The string was free to move during the first 1000 $\tau$ of the simulation and then was fixed at its converged position to collect statistics for the path.    
An equilibrium re-weighting procedure was applied during the first 1500 $\tau$  of the simulation every 10 $\tau$.
An additional 3000 $\tau$ steps of the WE method were simulated with the converged string to collect statistics.
The total simulation time was 4500 $\tau$.

The free energy associated with the path defined by the converged string is shown in Fig.~\ref{fig:ntrc_pmf}.
The free energy in each Voronoi bin, $G_{\alpha}$, is calculated from the WE simulation directly as $G_{\alpha} = -k_{B}T ln\left(\overline{w}_{\alpha}\right)$, where $\overline{w}_{\alpha}$ is the average weight assigned to bin $\alpha$.
Statistical errors for averages of the timeseries data used to calculate $\overline{w}_{\alpha}$ were estimated by autocorrelation analysis\cite{Chodera2007Use-of-the-weighted-} using the {\tt timeseries} module of the pyMBAR code\cite{Shirts2008Statistically-optima}.
The free energy along the string is shown in Fig.~\ref{fig:ntrc_pmf} and agrees with the barrier height and relative stability between $A$ and $B$ to within 1 kcal/mol of the results calculated from the FTS method (Figure 11 in Ref.~\onlinecite{Vanden-Eijnden2009Revisiting-the-finit}).

While we extended the simulation 3000 $\tau$ beyond the phase when re-weighting was applied, an accurate estimate of the free energy difference between $A$ and $B$ (within the 95 \% confidence interval) was obtained during the first 500 $\tau$ of the simulation.
During this same period the barrier height differed from the converged value by less than 1.3 kcal/mol.

While the free energy shows a distinct barrier along the path separating the inactive and active states, it is important to determine whether the converged path is dynamically relevant, and if the barrier corresponds to a mechanistic transition state.
To this end, we calculate the committor probability\cite{Bolhuis2002Transition-path,Du1998On-the-transiti,Geissler1999Kinetic-pathways-of-} for each bin, $\alpha$, along the string, $q^+_{\alpha}$, defined as the probability that fleeting trajectories initiated from that bin reach state $B$ before state $A$.
If the string accurately describes the dynamical reaction pathway, $q^+_{\alpha}=1/2$ should coincide with the barrier with neighboring bins rapidly asymptoting to zero and unity on either side of the barrier. 

For each bin, we select 500 random conformations drawn from the final 2500 $\tau$ iterations of the WE simulation.
A hundred conventional simulations are initiated from each conformation with initial velocities drawn randomly from a Boltzmann distribution at the same temperature as the WE simulations.
These simulations are propagated until the trajectory reaches state $A$ or $B$, and the terminating state is recorded.
A trajectory terminates in state $A$ ($B$) when its RMSD is $< 2$ ${\angstrom}$ from the reference conformation of $A$ ($B$) and $>3$ ${\angstrom}$ from $B$ ($A$).

Alternatively, the committor probability along the string can also be calculated directly from transition statistics gathered during the WE simulations by solving the following system of equations\cite{Noe2009Constructing-th}
\begin{eqnarray}
-q_i^{+} + \sum_{k{\in}I}T_{ik}q_k^{+} = -\sum_{k{\in}B}T_{ik} \text{ for  $i{\in}I$},
\label{eq:committor}
\end{eqnarray}
where $q_{i}^{+}$ is the probability that from bin $i$ the system will reach state $B$ before returning to state $A$, and $T_{i,j}$ is the probability of going from bin $i$ to $j$, which is calculated directly from the WE simulation.
Intermediate states, $I$, are those states which are not in $A$ or $B$. 

Committor probabilities calculated from both methods are shown in Fig.~\ref{fig:ntrc_committor} and are nearly identical over all of the bins.
The $q^+_{\alpha}=1/2$ bins coincide with the peak of the barrier in Fig.~\ref{fig:ntrc_pmf} between $\alpha=18$ and $19$.
Representative conformations from these bins are shown in the center panel of Fig.~\ref{fig:ntrc_struct}.
Taken together with the observation that the underlying distributions for each bin, $P\left(q^+_{\alpha}\right)$, are unimodal, the converged string appears to describe the reactive pathway between the inactive and actives states for this model of NtrC$^{r}$.

\section{Discussion}

We have introduced a variant of the WE method to sample transitions along a one-dimensional path embedded in the space of an arbitrary number of order parameters.
As in the finite-temperature string method\cite{Vanden-Eijnden2009Revisiting-the-finit} on which our method is based, an equidistantly spaced set of images along the path define a set of Voronoi cells that are analogous to the subdivision of phase space into bins in earlier versions of the WE method.
The images adaptively evolve toward the average weighted position of replicas that migrate through each bin during the WE simulation, subject to a smoothing reparameterization. 
The use of a dynamic set of bins that change in time during a WE simulation had been previously suggested\cite{Huber1996Weighted-ensemble-Br,Zhang2007Efficient-and-verifi,Zhang2010The-weighted-ensembl}, and here we demonstrate its practical use to confine sampling to the region of order parameter space along the transition pathway. 

Unlike in variants of the string method where the string images are used to either initiate replicas of the system\cite{Johnson2012Characterization-of-,Pan2008Finding-transition-p}, or to perform constrained sampling along a hyperplane coincident with string\cite{Weinan2005Finite-temperature-s}, here the path only serves to partition phase space.
Its utility when combined with the adaptive update scheme is that it allows the WE method to enhance sampling along the transition tube between states of interest.
In doing so, the string and associated Voronoi bins do not perturb the natural dynamics of the system.

The path of the string does, however, play an important role in determining the efficiency of the WE method.
Partitioning progress coordinate space along a pseudo one-dimensional curve converts the computation from one where the cost scales exponentially with the number of order parameters to one which is linear in the number of images along the string.
When a large number of progress coordinates are required to effectively sample a complex system, this change in the scaling behavior is likely to drastically reduce the computational cost of performing a WE simulation.
Additionally, so long as the string accurately approximates the dominant path of reactive flux, the majority of transitions along the string will occur between neighboring Voronoi bins. 
The reduced number of relevant bin interfaces across which transitions occur should decrease statistical errors in the rates estimates used in Eq.~\ref{eq:reweight}.
This likely increases the efficiency of the re-weighting procedure in Sec.~\ref{sec:reweighting} compared to its use with other binning regimes previously used with WE sampling.

The effectiveness of the string method as a way of partitioning phase space is subject to the assumption that the width of the transition tube about the string is small compared to its radius of curvature\cite{Ren2005Transition-pathways-}.
This assumption is a well-documented limitation of the string method that precludes its use for systems with multiple reactive channels or many important metastable states connected via a meshwork of isolated transition pathways.
The use of multiple strings to sample each pathway (using a modified version of the dual-direction scheme in Sec.~\ref{sec:drp}), or pairwise between well-defined states could circumvent this limitation, although accurately positing the presence and location of multiple pathways \emph{a priori} for complex systems in almost all cases is non-trivial.
   
\begin{figure}
\includegraphics{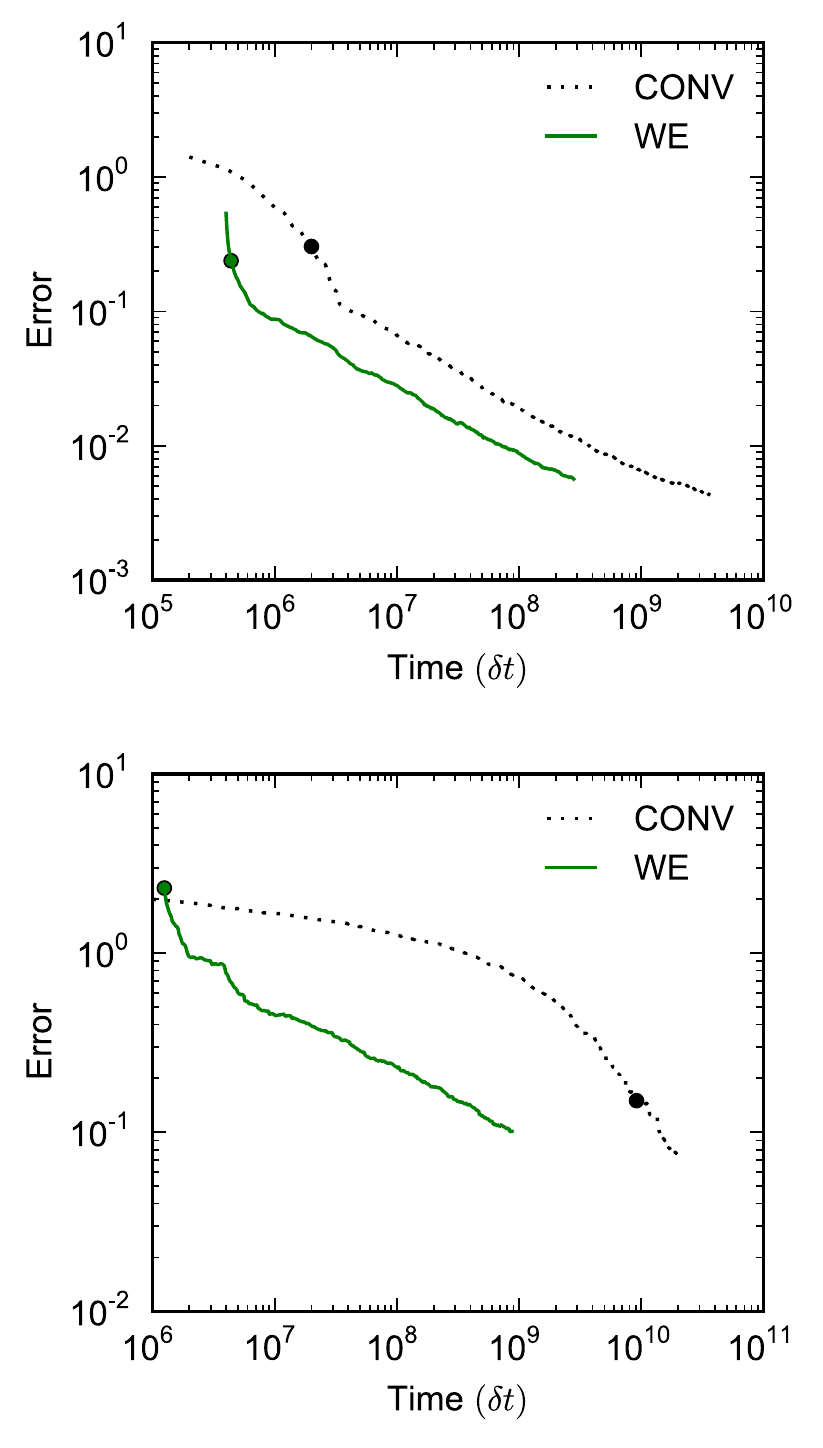}
\caption{
\label{fig:dpp_error}
Convergence of the steady-state distributions projected onto the $y$ axis for both conventional and WE simulations for the two-dimensional periodic potential where ${\alpha}=1.125$ (top) and ${\alpha}=2.25$ (bottom).
For both the conventional and WE simulations, the errors are averages over individual error curves calculated for ten independent simulations using Eqs.~(\ref{eq:dpp_error_a}) and (\ref{eq:dpp_error_b}).
The distributions for the WE simulations do not include statistics from the first 50$\tau$. 
The target distribution for the shallow potential (top) is calculated from a single long conventional simulation $4.0\times10^9$ steps in length. 
For the deep potential (bottom), the target distribution was obtained by averaging ten conventional simulations of $2.0\times10^{10}$ steps.
In each case, separate simulations were used to calculate the error and the target distributions. 
The solid circle marks the average time at which all histogram windows were populated for the ten simulations.
}
\end{figure}

\begin{figure*}
\includegraphics{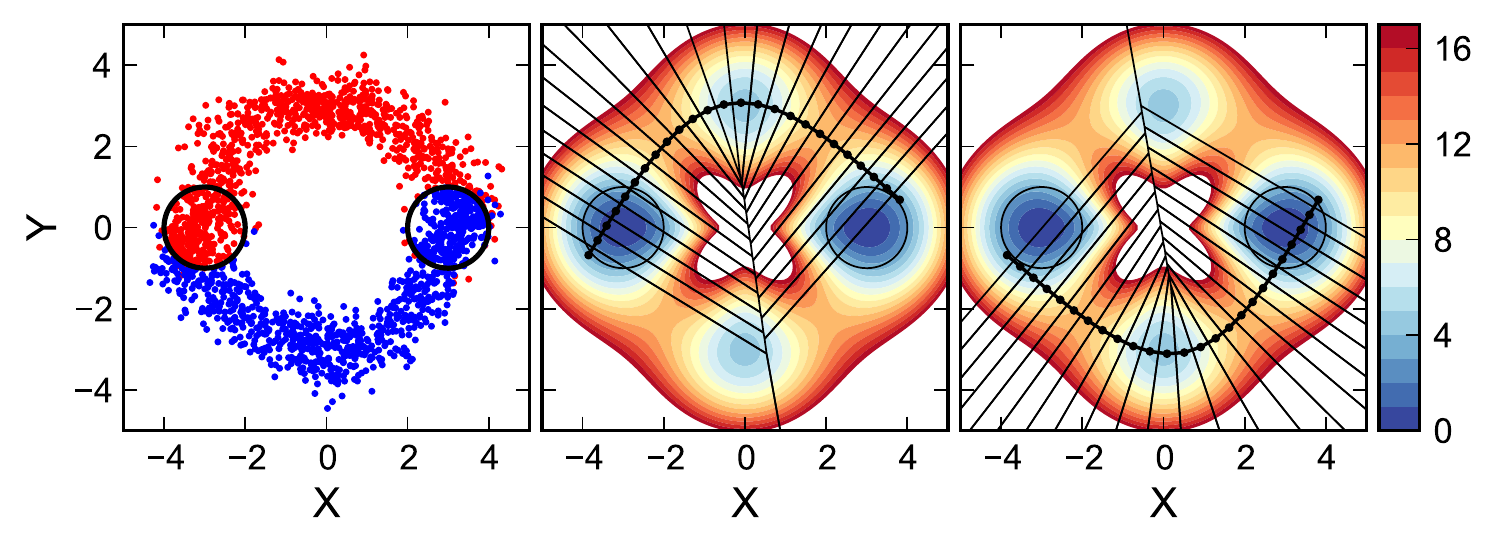}
\caption{
\label{fig:drp_potstr}
The two-dimensional ring potential with two pathways.
(left panel) An instantaneous snapshot of all of the active replicas during a representative iteration, where replicas that have last visited $A$ are shown in red, and those
that last visited $B$ shown in blue. 
The circles, centered at (-3,0) and (3,0), show the boundaries of the $A$ and $B$ states.
(center and right panels). The center and right panels show the converged strings with the corresponding Voronoi cells for the $A{\rightarrow}B$, and $B{\rightarrow}A$ transitions, respectively.
The circles delineate $A$ and $B$ as in the left panel.
Contour lines are separated by $kT$, and the corresponding color scale is show in the same units. }
\end{figure*}

\begin{figure}
\includegraphics{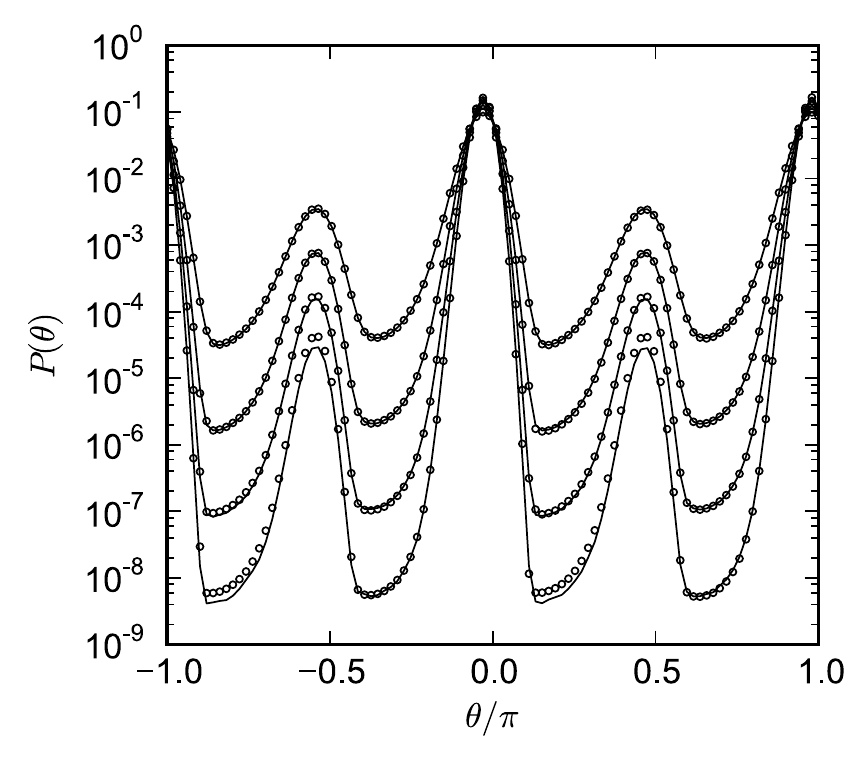}
\caption{
\label{fig:drp_distributions}
Projections of the steady-state distribution for the two-dimensional ring potential onto $\theta$ for ${\beta}$=1.0,1.5,2.0 and 2.5, obtained using conventional (lines) and weighted ensemble (circles) simulations.
The probability of finding a particle in either of the two metastable intermediates at $\theta=-\pi/2$ or $\pi/2$ decreases with increasing $\beta$. }
\end{figure}

\begin{figure}
\includegraphics{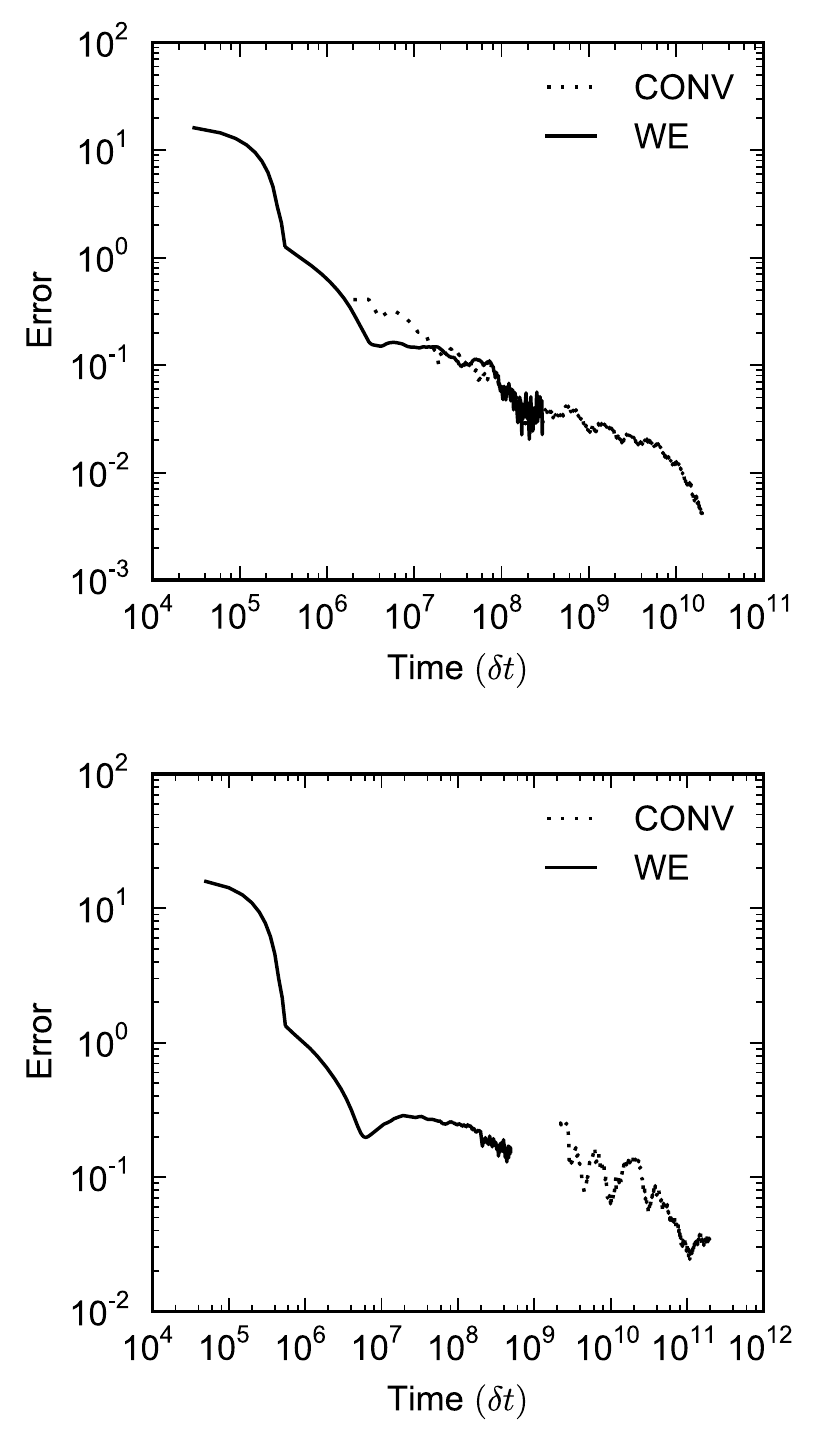}
\caption{
\label{fig:drp_error}
Convergence of the $A{\rightarrow}B$ rate constant for the two-dimensional ring potential where ${\beta}=1.5$ (top) and $2.5$ (bottom).
For both conventional and WE simulations, the errors are calculated using Eq.~\ref{eq:drp_error}, and are the averages of the error curves of ten independent simulations.
For the conventional simulations, an estimate of the error cannot be obtained until the first transition from $A$ to $B$ is observed.}
\end{figure}

\begin{figure}
\includegraphics{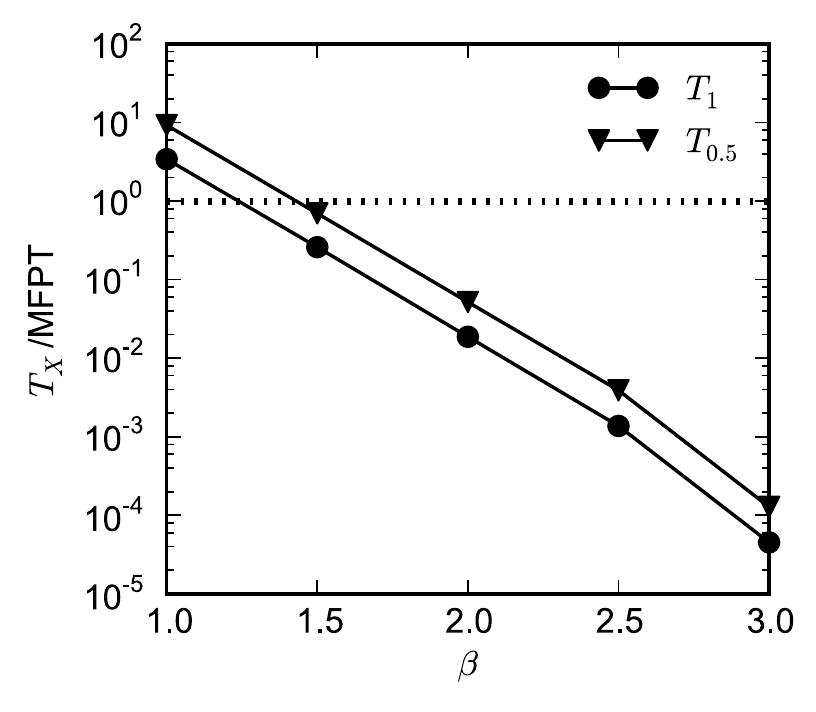}
\caption{Efficiency of the WE method for the two-dimensional ring potential as a function of inverse temperature, $\beta$.
Efficiency is defined as the ratio of the total aggregate simulation time, $T_X$, to the system's natural MFPT: $T_X/{\text{MFPT}}$, where $T_X$ is the time required to achieve a desired error, X.  
Thus, when the efficiency is 1 the total simulation time is the time required for the MFPT, which may be extremely long.
We carried out this analysis for an order of magnitude error estimate ($10^1$) of the rate $T_1$ and a factor of three error estimate ($10^{0.5} \sim 3$) of the rate $T_{0.5}$  
\label{fig:drp_rate_ord_mag}
}
\end{figure}

\begin{figure*}
\includegraphics{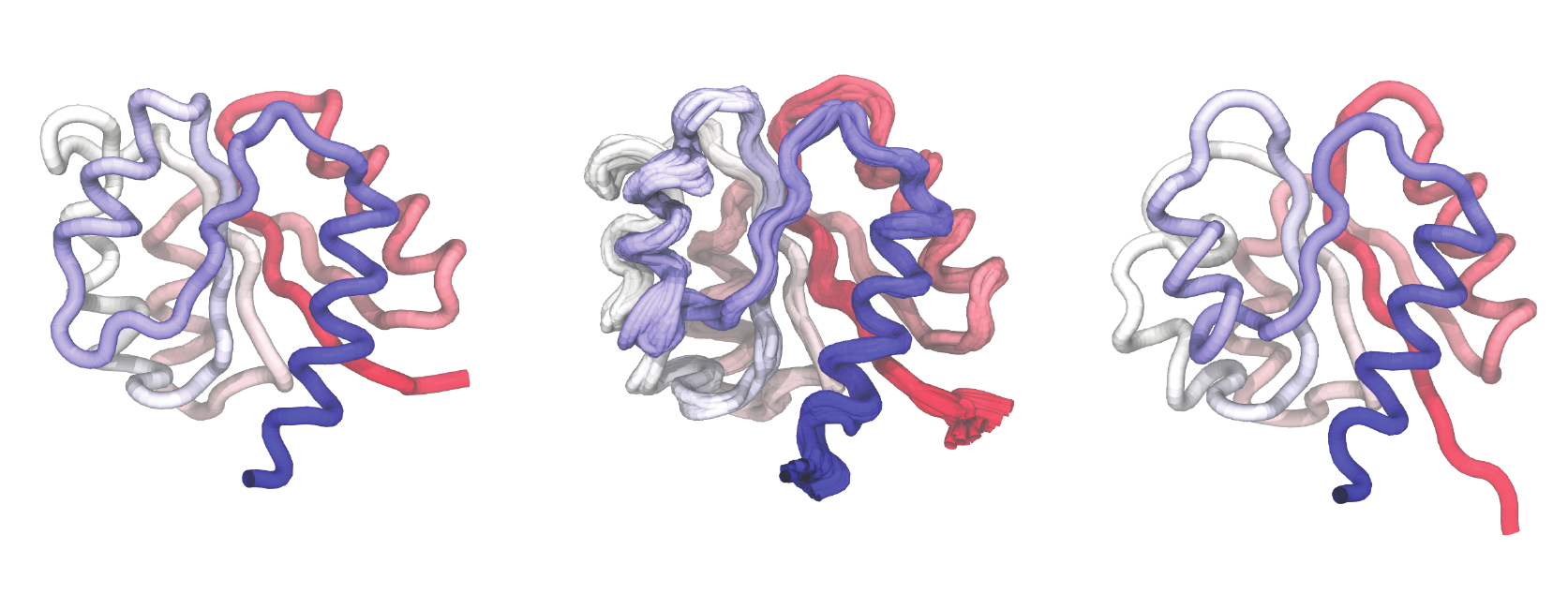}
\caption{The inactive (left) and active (right) conformations of the NtrC$^{r}$ receiver domain, generated from PDB IDs 1DC7 and 1DC8, respectively. 
(center) Ten conformations taken from the Voronoi bin corresponding most closely to the $q^+=0.5$ (bin index 18). 
Each conformation is the average of 5 randomly selected snapshots.
Coloring is based on residue indices starting from blue at the N-terminus and ending with red at the C-terminus.  
\label{fig:ntrc_struct}
}
\end{figure*}

\begin{figure}
\includegraphics{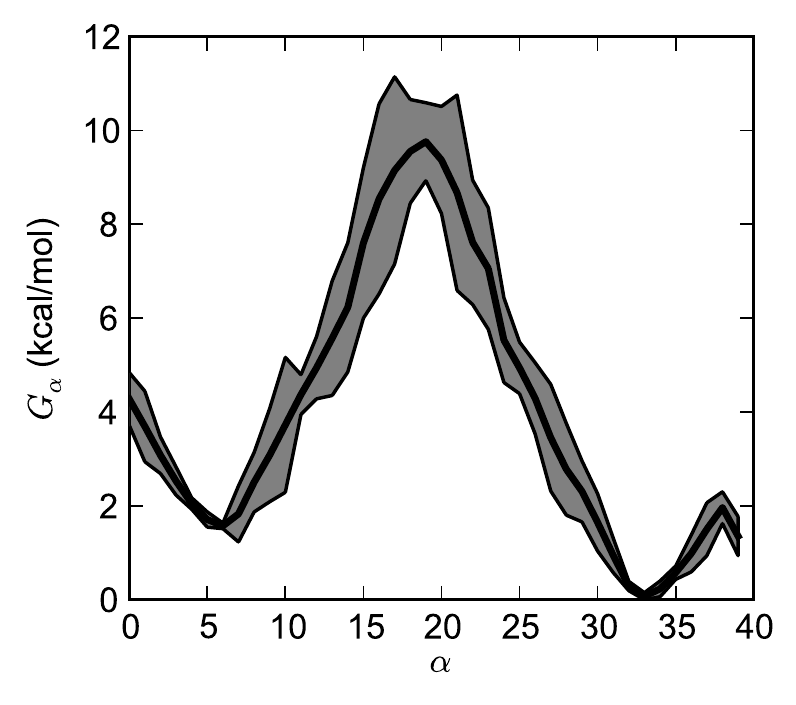}
\caption{Free energy $G_{\alpha}$ associated with each Voronoi tesselation along the string for the elastic network model of NtrC$^{r}$.
The dark shaded region denotes the 95\% conÞdence interval (two standard errors) for the free energy averaged over the last 3000 $\tau$ of the WE simulation.
\label{fig:ntrc_pmf}
}
\end{figure}

\begin{figure}
\includegraphics{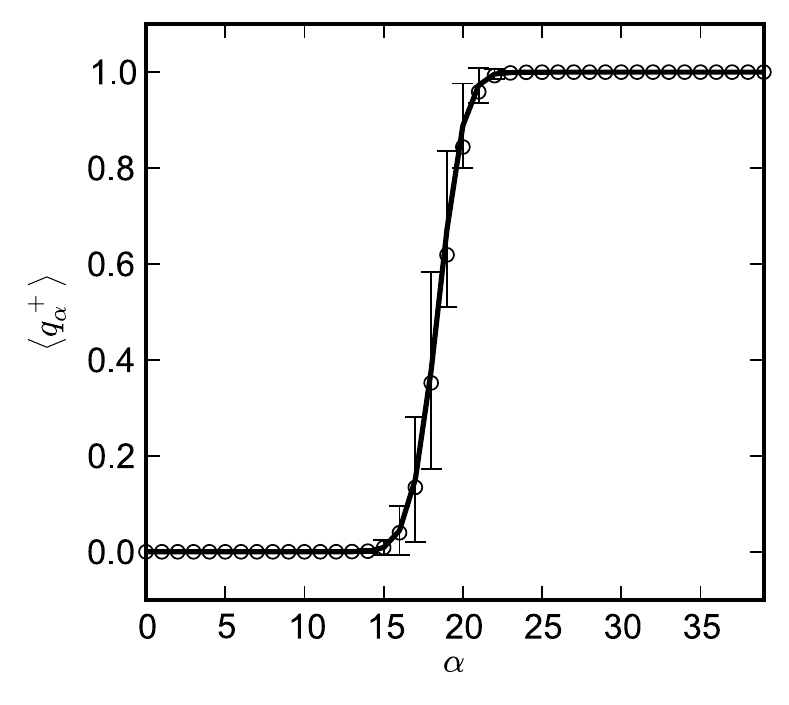}
\caption{Committor probability along the string for the elastic network model of NtrC$^{r}$.
The average committor probability of reaching state $B$ before state $A$, $q^+$, calculated from an ensemble of 500 conformations in each bin is shown as a solid line, with error bars equal to the standard deviation of the sample. 
Committors calculated from the WE simulation using Eq.~\ref{eq:committor} are shown as open circles.
\label{fig:ntrc_committor}
}
\end{figure}

In selecting the two driven 2D systems in sections \ref{sec:dpp} and \ref{sec:drp}, we sought to provide a direct comparison between our method and the NEUS-based string method of Dickson, \emph{et al}\cite{Dickson2009Nonequilibrium-umbre,Dickson2009Separating-forward-a}.
We carefully attempted to replicate the implementation of the underlying Brownian dynamics, as well as their convergence analysis to test the efficiency of the WE method.
Since the conventional dynamics in this study and the NEUS papers agree within statistical variation, the results presented in Figs.~\ref{fig:dpp_error},\ref{fig:drp_error} and \ref{fig:drp_rate_ord_mag} should be directly comparable. 
For both example systems, WE performs admirably and appears to show similar or better convergence characteristics than NEUS.
It is important to note, however, that in this comparison, the WE simulations use a higher density of replicas than in the NEUS studies.
In NEUS, a single replica is simulated per bin; in the WE simulations many replicas are run per bin, although each replica is run for a significantly shorter time than the replicas in the NEUS simulations per iteration.
We generally observe a super-linear gain in efficiency over some range of increasing the number of replicas and/or bins in the system (e.g. increasing the number of bins by a factor of 2 speeds convergence by $>2$ times).
Increasing the number of bins likely reduces the discretization errors associated with estimating the rates used in Eq.~\ref{eq:reweight} during the re-weighting step early in the simulation\cite{Prinz2011Markov-models-o}, and increasing the number of replicas reduces the variance in the inter-bin transition rates. 
Using multiple replicas per sampling region, which was suggested in a recent application of  the NEUS to the nonequilibirum folding and unfolding of coarse-grained model of RNA\cite{Dickson2011Flow-dependent-}, in concert with a finer discretization of phase space, may ameliorate the performance differences between the two methods.

Finally, we have presented a new string method based on sampling obtained via the Weighted Ensemble method that appears to have many advantages over previous methods. 
WE sampling is easy to implement both in terms of the data structures needed to track the replicas of the system and also by not requiring any special modification of the underlying dynamics to restrain replicas to a particular region of phase space, either through momentum reversal at a boundary\cite{Dickson2009Nonequilibrium-umbre,Dickson2009Separating-forward-a,Vanden-Eijnden2009Revisiting-the-finit} or soft-wall restraints\cite{Maragliano2009Free-energy-and}.
Additionally, it is not necessary to generate physical replicas along the initial string from the start state to the final state as it is in most other string methods. 
Often the initial string is generated using a simple linear interpolation, targeted molecular dynamics\cite{Pan2008Finding-transition-p} or with a coarse-grained model\cite{Zhu2010Pore-opening-and-clo}, all of which can lead to string images corresponding to unphysical intermediates for many systems of interest.
With the WE-based method, all replicas can be initiated in the start state and the string will evolve based on the natural dynamics of the system, without ever starting replicas based on these unphysical conformations.
In this regard, our method shares some similarities with the growing string method\cite{Peters2004A-growing-string-met}.
Despite this, some care should be taken when generating the initial path of the string.
The calculation could still become trapped if the initial path is separated from the dominant pathway by a large barrier orthogonal to the string.
Lastly, as we have shown here, the WE-based string method can be drastically more efficient than conventional sampling for calculating the rates and steady-state distributions for a range of equilibrium and nonequilibrium problems. 
Our method also outperforms the NEUS rare-event sampling method for two of the example systems studied here, but this result may depend on the system being simulated or it may be possible to tune the NEUS parameters to increase performance. 
This work lays the foundation for applying the WE-based string method to simulating rare transitions in more complex and realistic systems, especially when one or a small number of progress coordinates are insufficient to fully characterize the reaction coordinate.


%
%

%

\begin{acknowledgments}
We wish to thank Lillian Chong and Dan Zuckerman for critical reading of the manuscript, and Lillian Chong and Matt Zwier for their development of the WESTPA code's core functionality. 
We also thank Aaron Dinner and Alex Dickson for helpful discussions concerning the driven Brownian models used in the examples, and Eric Vanden-Eijnden and Maddalena Venturoli for sharing code related to the NtrC model.
This work was supported by National Institutes of Health (NIH) Grant No. R01-GM089740-01A1 (M.G.), NIH Grant No. T32-DK061296 (J.L.A) and FundScience (J.L.A).
\end{acknowledgments}


\begin{thebibliography}{57}%
\makeatletter
\providecommand \@ifxundefined [1]{%
 \@ifx{#1\undefined}
}%
\providecommand \@ifnum [1]{%
 \ifnum #1\expandafter \@firstoftwo
 \else \expandafter \@secondoftwo
 \fi
}%
\providecommand \@ifx [1]{%
 \ifx #1\expandafter \@firstoftwo
 \else \expandafter \@secondoftwo
 \fi
}%
\providecommand \natexlab [1]{#1}%
\providecommand \enquote  [1]{``#1''}%
\providecommand \bibnamefont  [1]{#1}%
\providecommand \bibfnamefont [1]{#1}%
\providecommand \citenamefont [1]{#1}%
\providecommand \href@noop [0]{\@secondoftwo}%
\providecommand \href [0]{\begingroup \@sanitize@url \@href}%
\providecommand \@href[1]{\@@startlink{#1}\@@href}%
\providecommand \@@href[1]{\endgroup#1\@@endlink}%
\providecommand \@sanitize@url [0]{\catcode `\\12\catcode `\$12\catcode
  `\&12\catcode `\#12\catcode `\^12\catcode `\_12\catcode `\%12\relax}%
\providecommand \@@startlink[1]{}%
\providecommand \@@endlink[0]{}%
\providecommand \url  [0]{\begingroup\@sanitize@url \@url }%
\providecommand \@url [1]{\endgroup\@href {#1}{\urlprefix }}%
\providecommand \urlprefix  [0]{URL }%
\providecommand \Eprint [0]{\href }%
\providecommand \doibase [0]{http://dx.doi.org/}%
\providecommand \selectlanguage [0]{\@gobble}%
\providecommand \bibinfo  [0]{\@secondoftwo}%
\providecommand \bibfield  [0]{\@secondoftwo}%
\providecommand \translation [1]{[#1]}%
\providecommand \BibitemOpen [0]{}%
\providecommand \bibitemStop [0]{}%
\providecommand \bibitemNoStop [0]{.\EOS\space}%
\providecommand \EOS [0]{\spacefactor3000\relax}%
\providecommand \BibitemShut  [1]{\csname bibitem#1\endcsname}%
\let\auto@bib@innerbib\@empty
\bibitem [{\citenamefont {Dellago}\ \emph {et~al.}(1998)\citenamefont
  {Dellago}, \citenamefont {Bolhuis}, \citenamefont {Csajka},\ and\
  \citenamefont {Chandler}}]{Dellago1998Transition-path}%
  \BibitemOpen
  \bibfield  {author} {\bibinfo {author} {\bibfnamefont {C.}~\bibnamefont
  {Dellago}}, \bibinfo {author} {\bibfnamefont {P.~G.}\ \bibnamefont
  {Bolhuis}}, \bibinfo {author} {\bibfnamefont {F.~S.}\ \bibnamefont {Csajka}},
  \ and\ \bibinfo {author} {\bibfnamefont {D.}~\bibnamefont {Chandler}},\
  }\bibfield  {title} {\enquote {\bibinfo {title} {Transition path sampling and
  the calculation of rate constants},}\ }\href {\doibase 10.1063/1.475562}
  {\bibfield  {journal} {\bibinfo  {journal} {The Journal of Chemical Physics}\
  }\textbf {\bibinfo {volume} {108}},\ \bibinfo {pages} {1964--1977} (\bibinfo
  {year} {1998})}\BibitemShut {NoStop}%
\bibitem [{\citenamefont {Woolf}(1998)}]{Woolf1998Path-corrected-}%
  \BibitemOpen
  \bibfield  {author} {\bibinfo {author} {\bibfnamefont {T.}~\bibnamefont
  {Woolf}},\ }\bibfield  {title} {\enquote {\bibinfo {title} {Path corrected
  functionals of stochastic trajectories: towards relative free energy and
  reaction coordinate calculations},}\ }\href@noop {} {\bibfield  {journal}
  {\bibinfo  {journal} {Chemical physics letters}\ }\textbf {\bibinfo {volume}
  {289}},\ \bibinfo {pages} {433--441} (\bibinfo {year} {1998})}\BibitemShut
  {NoStop}%
\bibitem [{\citenamefont {van Erp}, \citenamefont {Moroni},\ and\ \citenamefont
  {Bolhuis}(2003)}]{Erp2003A-novel-path-sa}%
  \BibitemOpen
  \bibfield  {author} {\bibinfo {author} {\bibfnamefont {T.~S.}\ \bibnamefont
  {van Erp}}, \bibinfo {author} {\bibfnamefont {D.}~\bibnamefont {Moroni}}, \
  and\ \bibinfo {author} {\bibfnamefont {P.~G.}\ \bibnamefont {Bolhuis}},\
  }\bibfield  {title} {\enquote {\bibinfo {title} {A novel path sampling method
  for the calculation of rate constants},}\ }\href {\doibase 10.1063/1.1562614}
  {\bibfield  {journal} {\bibinfo  {journal} {The Journal of Chemical Physics}\
  }\textbf {\bibinfo {volume} {118}},\ \bibinfo {pages} {7762--7774} (\bibinfo
  {year} {2003})}\BibitemShut {NoStop}%
\bibitem [{\citenamefont {Faradjian}\ and\ \citenamefont
  {Elber}(2004)}]{Faradjian2004Computing-time-}%
  \BibitemOpen
  \bibfield  {author} {\bibinfo {author} {\bibfnamefont {A.~K.}\ \bibnamefont
  {Faradjian}}\ and\ \bibinfo {author} {\bibfnamefont {R.}~\bibnamefont
  {Elber}},\ }\bibfield  {title} {\enquote {\bibinfo {title} {Computing time
  scales from reaction coordinates by milestoning},}\ }\href {\doibase
  10.1063/1.1738640} {\bibfield  {journal} {\bibinfo  {journal} {J Chem Phys}\
  }\textbf {\bibinfo {volume} {120}},\ \bibinfo {pages} {10880--9} (\bibinfo
  {year} {2004})}\BibitemShut {NoStop}%
\bibitem [{\citenamefont {Allen}, \citenamefont {Warren},\ and\ \citenamefont
  {Ten~Wolde}(2005)}]{Allen2005Sampling-rare-switch}%
  \BibitemOpen
  \bibfield  {author} {\bibinfo {author} {\bibfnamefont {R.~J.}\ \bibnamefont
  {Allen}}, \bibinfo {author} {\bibfnamefont {P.~B.}\ \bibnamefont {Warren}}, \
  and\ \bibinfo {author} {\bibfnamefont {P.~R.}\ \bibnamefont {Ten~Wolde}},\
  }\bibfield  {title} {\enquote {\bibinfo {title} {Sampling rare switching
  events in biochemical networks},}\ }\href@noop {} {\bibfield  {journal}
  {\bibinfo  {journal} {Phys Rev Lett}\ }\textbf {\bibinfo {volume} {94}},\
  \bibinfo {pages} {018104} (\bibinfo {year} {2005})}\BibitemShut {NoStop}%
\bibitem [{\citenamefont {Warmflash}, \citenamefont {Bhimalapuram},\ and\
  \citenamefont {Dinner}(2007)}]{Warmflash2007Umbrella-sampli}%
  \BibitemOpen
  \bibfield  {author} {\bibinfo {author} {\bibfnamefont {A.}~\bibnamefont
  {Warmflash}}, \bibinfo {author} {\bibfnamefont {P.}~\bibnamefont
  {Bhimalapuram}}, \ and\ \bibinfo {author} {\bibfnamefont {A.~R.}\
  \bibnamefont {Dinner}},\ }\bibfield  {title} {\enquote {\bibinfo {title}
  {Umbrella sampling for nonequilibrium processes},}\ }\href {\doibase
  10.1063/1.2784118} {\bibfield  {journal} {\bibinfo  {journal} {J Chem Phys}\
  }\textbf {\bibinfo {volume} {127}},\ \bibinfo {pages} {154112} (\bibinfo
  {year} {2007})}\BibitemShut {NoStop}%
\bibitem [{\citenamefont {Bolhuis}\ \emph {et~al.}(2002)\citenamefont
  {Bolhuis}, \citenamefont {Chandler}, \citenamefont {Dellago},\ and\
  \citenamefont {Geissler}}]{Bolhuis2002Transition-path}%
  \BibitemOpen
  \bibfield  {author} {\bibinfo {author} {\bibfnamefont {P.~G.}\ \bibnamefont
  {Bolhuis}}, \bibinfo {author} {\bibfnamefont {D.}~\bibnamefont {Chandler}},
  \bibinfo {author} {\bibfnamefont {C.}~\bibnamefont {Dellago}}, \ and\
  \bibinfo {author} {\bibfnamefont {P.~L.}\ \bibnamefont {Geissler}},\
  }\bibfield  {title} {\enquote {\bibinfo {title} {Transition path sampling:
  throwing ropes over rough mountain passes, in the dark},}\ }\href {\doibase
  10.1146/annurev.physchem.53.082301.113146} {\bibfield  {journal} {\bibinfo
  {journal} {Annu Rev Phys Chem}\ }\textbf {\bibinfo {volume} {53}},\ \bibinfo
  {pages} {291--318} (\bibinfo {year} {2002})}\BibitemShut {NoStop}%
\bibitem [{\citenamefont {Dellago}\ and\ \citenamefont
  {Bolhuis}(2009)}]{Dellago2009Transition-path}%
  \BibitemOpen
  \bibfield  {author} {\bibinfo {author} {\bibfnamefont {C.}~\bibnamefont
  {Dellago}}\ and\ \bibinfo {author} {\bibfnamefont {P.}~\bibnamefont
  {Bolhuis}},\ }\bibfield  {title} {\enquote {\bibinfo {title} {Transition path
  sampling and other advanced simulation techniques for rare events},}\
  }\href@noop {} {\bibfield  {journal} {\bibinfo  {journal} {Advanced computer
  simulation approaches for soft matter sciences III}\ ,\ \bibinfo {pages}
  {167--233}} (\bibinfo {year} {2009})}\BibitemShut {NoStop}%
\bibitem [{\citenamefont {Zwier}\ and\ \citenamefont
  {Chong}(2010)}]{Zwier2010Reaching-biolog}%
  \BibitemOpen
  \bibfield  {author} {\bibinfo {author} {\bibfnamefont {M.~C.}\ \bibnamefont
  {Zwier}}\ and\ \bibinfo {author} {\bibfnamefont {L.~T.}\ \bibnamefont
  {Chong}},\ }\bibfield  {title} {\enquote {\bibinfo {title} {Reaching
  biological timescales with all-atom molecular dynamics simulations},}\ }\href
  {\doibase 10.1016/j.coph.2010.09.008} {\bibfield  {journal} {\bibinfo
  {journal} {Curr Opin Pharmacol}\ }\textbf {\bibinfo {volume} {10}},\ \bibinfo
  {pages} {745--52} (\bibinfo {year} {2010})}\BibitemShut {NoStop}%
\bibitem [{\citenamefont {Elber}\ and\ \citenamefont
  {Karplus}(1987)}]{Elber1987A-method-for-determi}%
  \BibitemOpen
  \bibfield  {author} {\bibinfo {author} {\bibfnamefont {R.}~\bibnamefont
  {Elber}}\ and\ \bibinfo {author} {\bibfnamefont {M.}~\bibnamefont
  {Karplus}},\ }\bibfield  {title} {\enquote {\bibinfo {title} {A method for
  determining reaction paths in large molecules: application to myoglobin},}\
  }\href@noop {} {\bibfield  {journal} {\bibinfo  {journal} {Chemical Physics
  Letters}\ }\textbf {\bibinfo {volume} {139}},\ \bibinfo {pages} {375--380}
  (\bibinfo {year} {1987})}\BibitemShut {NoStop}%
\bibitem [{\citenamefont {Fischer}\ and\ \citenamefont
  {Karplus}(1992)}]{Fischer1992Conjugate-peak-}%
  \BibitemOpen
  \bibfield  {author} {\bibinfo {author} {\bibfnamefont {S.}~\bibnamefont
  {Fischer}}\ and\ \bibinfo {author} {\bibfnamefont {M.}~\bibnamefont
  {Karplus}},\ }\bibfield  {title} {\enquote {\bibinfo {title} {Conjugate peak
  refinement: an algorithm for finding reaction paths and accurate transition
  states in systems with many degrees of freedom},}\ }\href@noop {} {\bibfield
  {journal} {\bibinfo  {journal} {Chemical physics letters}\ }\textbf {\bibinfo
  {volume} {194}},\ \bibinfo {pages} {252--261} (\bibinfo {year}
  {1992})}\BibitemShut {NoStop}%
\bibitem [{\citenamefont {Henkelman}\ and\ \citenamefont
  {J{\'o}nsson}(2000)}]{Henkelman2000Improved-tangen}%
  \BibitemOpen
  \bibfield  {author} {\bibinfo {author} {\bibfnamefont {G.}~\bibnamefont
  {Henkelman}}\ and\ \bibinfo {author} {\bibfnamefont {H.}~\bibnamefont
  {J{\'o}nsson}},\ }\bibfield  {title} {\enquote {\bibinfo {title} {Improved
  tangent estimate in the nudged elastic band method for finding minimum energy
  paths and saddle points},}\ }\href@noop {} {\bibfield  {journal} {\bibinfo
  {journal} {The Journal of Chemical Physics}\ }\textbf {\bibinfo {volume}
  {113}},\ \bibinfo {pages} {9978--9985} (\bibinfo {year} {2000})}\BibitemShut
  {NoStop}%
\bibitem [{\citenamefont {Weinan}, \citenamefont {Weiqing},\ and\ \citenamefont
  {Vanden-Eijnden}(2002)}]{Weinan2002String-method-f}%
  \BibitemOpen
  \bibfield  {author} {\bibinfo {author} {\bibfnamefont {E.}~\bibnamefont
  {Weinan}}, \bibinfo {author} {\bibfnamefont {R.}~\bibnamefont {Weiqing}}, \
  and\ \bibinfo {author} {\bibfnamefont {E.}~\bibnamefont {Vanden-Eijnden}},\
  }\bibfield  {title} {\enquote {\bibinfo {title} {String method for the study
  of rare events},}\ }\href@noop {} {\bibfield  {journal} {\bibinfo  {journal}
  {Physical review B. Condensed matter and materials physics}\ }\textbf
  {\bibinfo {volume} {66}},\ \bibinfo {pages} {052301--1} (\bibinfo {year}
  {2002})}\BibitemShut {NoStop}%
\bibitem [{\citenamefont {Maragliano}\ \emph {et~al.}(2006)\citenamefont
  {Maragliano}, \citenamefont {Fischer}, \citenamefont {Vanden-Eijnden},\ and\
  \citenamefont {Ciccotti}}]{Maragliano2006String-method-i}%
  \BibitemOpen
  \bibfield  {author} {\bibinfo {author} {\bibfnamefont {L.}~\bibnamefont
  {Maragliano}}, \bibinfo {author} {\bibfnamefont {A.}~\bibnamefont {Fischer}},
  \bibinfo {author} {\bibfnamefont {E.}~\bibnamefont {Vanden-Eijnden}}, \ and\
  \bibinfo {author} {\bibfnamefont {G.}~\bibnamefont {Ciccotti}},\ }\bibfield
  {title} {\enquote {\bibinfo {title} {String method in collective variables:
  minimum free energy paths and isocommittor surfaces},}\ }\href {\doibase
  10.1063/1.2212942} {\bibfield  {journal} {\bibinfo  {journal} {J Chem Phys}\
  }\textbf {\bibinfo {volume} {125}},\ \bibinfo {pages} {24106} (\bibinfo
  {year} {2006})}\BibitemShut {NoStop}%
\bibitem [{\citenamefont {Weinan}, \citenamefont {Ren},\ and\ \citenamefont
  {Vanden-Eijnden}(2005)}]{Weinan2005Finite-temperature-s}%
  \BibitemOpen
  \bibfield  {author} {\bibinfo {author} {\bibfnamefont {E.}~\bibnamefont
  {Weinan}}, \bibinfo {author} {\bibfnamefont {W.}~\bibnamefont {Ren}}, \ and\
  \bibinfo {author} {\bibfnamefont {E.}~\bibnamefont {Vanden-Eijnden}},\
  }\bibfield  {title} {\enquote {\bibinfo {title} {Finite temperature string
  method for the study of rare events},}\ }\href@noop {} {\bibfield  {journal}
  {\bibinfo  {journal} {The Journal of Physical Chemistry B}\ }\textbf
  {\bibinfo {volume} {109}},\ \bibinfo {pages} {6688--6693} (\bibinfo {year}
  {2005})}\BibitemShut {NoStop}%
\bibitem [{\citenamefont {Pan}, \citenamefont {Sezer},\ and\ \citenamefont
  {Roux}(2008)}]{Pan2008Finding-transition-p}%
  \BibitemOpen
  \bibfield  {author} {\bibinfo {author} {\bibfnamefont {A.~C.}\ \bibnamefont
  {Pan}}, \bibinfo {author} {\bibfnamefont {D.}~\bibnamefont {Sezer}}, \ and\
  \bibinfo {author} {\bibfnamefont {B.}~\bibnamefont {Roux}},\ }\bibfield
  {title} {\enquote {\bibinfo {title} {Finding transition pathways using the
  string method with swarms of trajectories},}\ }\href {\doibase
  10.1021/jp0777059} {\bibfield  {journal} {\bibinfo  {journal} {J Phys Chem
  B}\ }\textbf {\bibinfo {volume} {112}},\ \bibinfo {pages} {3432--40}
  (\bibinfo {year} {2008})}\BibitemShut {NoStop}%
\bibitem [{\citenamefont {Pan}\ and\ \citenamefont
  {Roux}(2008)}]{Pan2008Building-Markov}%
  \BibitemOpen
  \bibfield  {author} {\bibinfo {author} {\bibfnamefont {A.~C.}\ \bibnamefont
  {Pan}}\ and\ \bibinfo {author} {\bibfnamefont {B.}~\bibnamefont {Roux}},\
  }\bibfield  {title} {\enquote {\bibinfo {title} {Building markov state models
  along pathways to determine free energies and rates of transitions},}\ }\href
  {\doibase 10.1063/1.2959573} {\bibfield  {journal} {\bibinfo  {journal} {J
  Chem Phys}\ }\textbf {\bibinfo {volume} {129}},\ \bibinfo {pages} {064107}
  (\bibinfo {year} {2008})}\BibitemShut {NoStop}%
\bibitem [{\citenamefont {Vanden-Eijnden}\ and\ \citenamefont
  {Venturoli}(2009{\natexlab{a}})}]{Vanden-Eijnden2009Revisiting-the-finit}%
  \BibitemOpen
  \bibfield  {author} {\bibinfo {author} {\bibfnamefont {E.}~\bibnamefont
  {Vanden-Eijnden}}\ and\ \bibinfo {author} {\bibfnamefont {M.}~\bibnamefont
  {Venturoli}},\ }\bibfield  {title} {\enquote {\bibinfo {title} {Revisiting
  the finite temperature string method for the calculation of reaction tubes
  and free energies},}\ }\href@noop {} {\bibfield  {journal} {\bibinfo
  {journal} {The Journal of chemical physics}\ }\textbf {\bibinfo {volume}
  {130}},\ \bibinfo {pages} {194103} (\bibinfo {year}
  {2009}{\natexlab{a}})}\BibitemShut {NoStop}%
\bibitem [{\citenamefont {Dickson}, \citenamefont {Warmflash},\ and\
  \citenamefont
  {Dinner}(2009{\natexlab{a}})}]{Dickson2009Nonequilibrium-umbre}%
  \BibitemOpen
  \bibfield  {author} {\bibinfo {author} {\bibfnamefont {A.}~\bibnamefont
  {Dickson}}, \bibinfo {author} {\bibfnamefont {A.}~\bibnamefont {Warmflash}},
  \ and\ \bibinfo {author} {\bibfnamefont {A.~R.}\ \bibnamefont {Dinner}},\
  }\bibfield  {title} {\enquote {\bibinfo {title} {Nonequilibrium umbrella
  sampling in spaces of many order parameters},}\ }\href {\doibase
  10.1063/1.3070677} {\bibfield  {journal} {\bibinfo  {journal} {J Chem Phys}\
  }\textbf {\bibinfo {volume} {130}},\ \bibinfo {pages} {074104} (\bibinfo
  {year} {2009}{\natexlab{a}})}\BibitemShut {NoStop}%
\bibitem [{\citenamefont {Rogal}\ \emph {et~al.}(2010)\citenamefont {Rogal},
  \citenamefont {Lechner}, \citenamefont {Juraszek}, \citenamefont {Ensing},\
  and\ \citenamefont {Bolhuis}}]{Rogal2010The-reweighted-path-}%
  \BibitemOpen
  \bibfield  {author} {\bibinfo {author} {\bibfnamefont {J.}~\bibnamefont
  {Rogal}}, \bibinfo {author} {\bibfnamefont {W.}~\bibnamefont {Lechner}},
  \bibinfo {author} {\bibfnamefont {J.}~\bibnamefont {Juraszek}}, \bibinfo
  {author} {\bibfnamefont {B.}~\bibnamefont {Ensing}}, \ and\ \bibinfo {author}
  {\bibfnamefont {P.~G.}\ \bibnamefont {Bolhuis}},\ }\bibfield  {title}
  {\enquote {\bibinfo {title} {The reweighted path ensemble},}\ }\href
  {\doibase 10.1063/1.3491817} {\bibfield  {journal} {\bibinfo  {journal} {J
  Chem Phys}\ }\textbf {\bibinfo {volume} {133}},\ \bibinfo {pages} {174109}
  (\bibinfo {year} {2010})}\BibitemShut {NoStop}%
\bibitem [{\citenamefont {Johnson}\ and\ \citenamefont
  {Hummer}(2012)}]{Johnson2012Characterization-of-}%
  \BibitemOpen
  \bibfield  {author} {\bibinfo {author} {\bibfnamefont {M.~E.}\ \bibnamefont
  {Johnson}}\ and\ \bibinfo {author} {\bibfnamefont {G.}~\bibnamefont
  {Hummer}},\ }\bibfield  {title} {\enquote {\bibinfo {title} {Characterization
  of a dynamic string method for the construction of transition pathways in
  molecular reactions},}\ }\href {\doibase 10.1021/jp212611k} {\bibfield
  {journal} {\bibinfo  {journal} {J Phys Chem B}\ }\textbf {\bibinfo {volume}
  {116}},\ \bibinfo {pages} {8573--83} (\bibinfo {year} {2012})}\BibitemShut
  {NoStop}%
\bibitem [{\citenamefont {Dickson}, \citenamefont {Huang},\ and\ \citenamefont
  {Post}(2012)}]{Dickson2012Unrestrained-Co}%
  \BibitemOpen
  \bibfield  {author} {\bibinfo {author} {\bibfnamefont {B.~M.}\ \bibnamefont
  {Dickson}}, \bibinfo {author} {\bibfnamefont {H.}~\bibnamefont {Huang}}, \
  and\ \bibinfo {author} {\bibfnamefont {C.~B.}\ \bibnamefont {Post}},\
  }\bibfield  {title} {\enquote {\bibinfo {title} {Unrestrained computation of
  free energy along a path},}\ }\href {\doibase 10.1021/jp304720m} {\bibfield
  {journal} {\bibinfo  {journal} {J Phys Chem B}\ } (\bibinfo {year} {2012}),\
  10.1021/jp304720m}\BibitemShut {NoStop}%
\bibitem [{\citenamefont {Ren}\ \emph {et~al.}(2005)\citenamefont {Ren},
  \citenamefont {Vanden-Eijnden}, \citenamefont {Maragakis},\ and\
  \citenamefont {Weinan}}]{Ren2005Transition-pathways-}%
  \BibitemOpen
  \bibfield  {author} {\bibinfo {author} {\bibfnamefont {W.}~\bibnamefont
  {Ren}}, \bibinfo {author} {\bibfnamefont {E.}~\bibnamefont {Vanden-Eijnden}},
  \bibinfo {author} {\bibfnamefont {P.}~\bibnamefont {Maragakis}}, \ and\
  \bibinfo {author} {\bibfnamefont {E.}~\bibnamefont {Weinan}},\ }\bibfield
  {title} {\enquote {\bibinfo {title} {Transition pathways in complex systems:
  Application of the finite-temperature string method to the alanine
  dipeptide},}\ }\href@noop {} {\bibfield  {journal} {\bibinfo  {journal} {The
  Journal of chemical physics}\ }\textbf {\bibinfo {volume} {123}},\ \bibinfo
  {pages} {134109} (\bibinfo {year} {2005})}\BibitemShut {NoStop}%
\bibitem [{\citenamefont {Miller}, \citenamefont {Vanden-Eijnden},\ and\
  \citenamefont {Chandler}(2007)}]{Miller2007Solvent-coarse-}%
  \BibitemOpen
  \bibfield  {author} {\bibinfo {author} {\bibfnamefont {T.~F.}\ \bibnamefont
  {Miller}, \bibfnamefont {3rd}}, \bibinfo {author} {\bibfnamefont
  {E.}~\bibnamefont {Vanden-Eijnden}}, \ and\ \bibinfo {author} {\bibfnamefont
  {D.}~\bibnamefont {Chandler}},\ }\bibfield  {title} {\enquote {\bibinfo
  {title} {Solvent coarse-graining and the string method applied to the
  hydrophobic collapse of a hydrated chain},}\ }\href {\doibase
  10.1073/pnas.0705830104} {\bibfield  {journal} {\bibinfo  {journal} {Proc
  Natl Acad Sci U S A}\ }\textbf {\bibinfo {volume} {104}},\ \bibinfo {pages}
  {14559--64} (\bibinfo {year} {2007})}\BibitemShut {NoStop}%
\bibitem [{\citenamefont {Ovchinnikov}\ \emph {et~al.}(2011)\citenamefont
  {Ovchinnikov}, \citenamefont {Cecchini}, \citenamefont {Vanden-Eijnden},\
  and\ \citenamefont {Karplus}}]{Ovchinnikov2011A-conformational-tra}%
  \BibitemOpen
  \bibfield  {author} {\bibinfo {author} {\bibfnamefont {V.}~\bibnamefont
  {Ovchinnikov}}, \bibinfo {author} {\bibfnamefont {M.}~\bibnamefont
  {Cecchini}}, \bibinfo {author} {\bibfnamefont {E.}~\bibnamefont
  {Vanden-Eijnden}}, \ and\ \bibinfo {author} {\bibfnamefont {M.}~\bibnamefont
  {Karplus}},\ }\bibfield  {title} {\enquote {\bibinfo {title} {A
  conformational transition in the myosin {VI} converter contributes to the
  variable step size},}\ }\href {\doibase 10.1016/j.bpj.2011.09.044} {\bibfield
   {journal} {\bibinfo  {journal} {Biophys J}\ }\textbf {\bibinfo {volume}
  {101}},\ \bibinfo {pages} {2436--44} (\bibinfo {year} {2011})}\BibitemShut
  {NoStop}%
\bibitem [{\citenamefont {Bhatt}\ and\ \citenamefont
  {Zuckerman}(2010)}]{Bhatt2010Heterogeneous-p}%
  \BibitemOpen
  \bibfield  {author} {\bibinfo {author} {\bibfnamefont {D.}~\bibnamefont
  {Bhatt}}\ and\ \bibinfo {author} {\bibfnamefont {D.~M.}\ \bibnamefont
  {Zuckerman}},\ }\bibfield  {title} {\enquote {\bibinfo {title} {Heterogeneous
  path ensembles for conformational transitions in semi-atomistic models of
  adenylate kinase},}\ }\href {\doibase 10.1021/ct100406t} {\bibfield
  {journal} {\bibinfo  {journal} {J Chem Theory Comput}\ }\textbf {\bibinfo
  {volume} {6}},\ \bibinfo {pages} {3527--3539} (\bibinfo {year}
  {2010})}\BibitemShut {NoStop}%
\bibitem [{\citenamefont {Adelman}\ \emph {et~al.}(2011)\citenamefont
  {Adelman}, \citenamefont {Dale}, \citenamefont {Zwier}, \citenamefont
  {Bhatt}, \citenamefont {Chong}, \citenamefont {Zuckerman},\ and\
  \citenamefont {Grabe}}]{Adelman2011Simulations-of-the-a}%
  \BibitemOpen
  \bibfield  {author} {\bibinfo {author} {\bibfnamefont {J.~L.}\ \bibnamefont
  {Adelman}}, \bibinfo {author} {\bibfnamefont {A.~L.}\ \bibnamefont {Dale}},
  \bibinfo {author} {\bibfnamefont {M.~C.}\ \bibnamefont {Zwier}}, \bibinfo
  {author} {\bibfnamefont {D.}~\bibnamefont {Bhatt}}, \bibinfo {author}
  {\bibfnamefont {L.~T.}\ \bibnamefont {Chong}}, \bibinfo {author}
  {\bibfnamefont {D.~M.}\ \bibnamefont {Zuckerman}}, \ and\ \bibinfo {author}
  {\bibfnamefont {M.}~\bibnamefont {Grabe}},\ }\bibfield  {title} {\enquote
  {\bibinfo {title} {Simulations of the alternating access mechanism of the
  sodium symporter {M}hp1},}\ }\href {\doibase 10.1016/j.bpj.2011.09.061}
  {\bibfield  {journal} {\bibinfo  {journal} {Biophys J}\ }\textbf {\bibinfo
  {volume} {101}},\ \bibinfo {pages} {2399--407} (\bibinfo {year}
  {2011})}\BibitemShut {NoStop}%
\bibitem [{\citenamefont {Huber}\ and\ \citenamefont
  {Kim}(1996)}]{Huber1996Weighted-ensemble-Br}%
  \BibitemOpen
  \bibfield  {author} {\bibinfo {author} {\bibfnamefont {G.~A.}\ \bibnamefont
  {Huber}}\ and\ \bibinfo {author} {\bibfnamefont {S.}~\bibnamefont {Kim}},\
  }\bibfield  {title} {\enquote {\bibinfo {title} {Weighted-ensemble brownian
  dynamics simulations for protein association reactions},}\ }\href {\doibase
  10.1016/S0006-3495(96)79552-8} {\bibfield  {journal} {\bibinfo  {journal}
  {Biophys J}\ }\textbf {\bibinfo {volume} {70}},\ \bibinfo {pages} {97--110}
  (\bibinfo {year} {1996})}\BibitemShut {NoStop}%
\bibitem [{\citenamefont {Zhang}, \citenamefont {Jasnow},\ and\ \citenamefont
  {Zuckerman}(2007)}]{Zhang2007Efficient-and-verifi}%
  \BibitemOpen
  \bibfield  {author} {\bibinfo {author} {\bibfnamefont {B.~W.}\ \bibnamefont
  {Zhang}}, \bibinfo {author} {\bibfnamefont {D.}~\bibnamefont {Jasnow}}, \
  and\ \bibinfo {author} {\bibfnamefont {D.~M.}\ \bibnamefont {Zuckerman}},\
  }\bibfield  {title} {\enquote {\bibinfo {title} {Efficient and verified
  simulation of a path ensemble for conformational change in a united-residue
  model of calmodulin},}\ }\href {\doibase 10.1073/pnas.0706349104} {\bibfield
  {journal} {\bibinfo  {journal} {Proc Natl Acad Sci U S A}\ }\textbf {\bibinfo
  {volume} {104}},\ \bibinfo {pages} {18043--8} (\bibinfo {year}
  {2007})}\BibitemShut {NoStop}%
\bibitem [{\citenamefont {Zhang}, \citenamefont {Jasnow},\ and\ \citenamefont
  {Zuckerman}(2010)}]{Zhang2010The-weighted-ensembl}%
  \BibitemOpen
  \bibfield  {author} {\bibinfo {author} {\bibfnamefont {B.~W.}\ \bibnamefont
  {Zhang}}, \bibinfo {author} {\bibfnamefont {D.}~\bibnamefont {Jasnow}}, \
  and\ \bibinfo {author} {\bibfnamefont {D.~M.}\ \bibnamefont {Zuckerman}},\
  }\bibfield  {title} {\enquote {\bibinfo {title} {The "weighted ensemble" path
  sampling method is statistically exact for a broad class of stochastic
  processes and binning procedures},}\ }\href {\doibase 10.1063/1.3306345}
  {\bibfield  {journal} {\bibinfo  {journal} {J Chem Phys}\ }\textbf {\bibinfo
  {volume} {132}},\ \bibinfo {pages} {054107} (\bibinfo {year}
  {2010})}\BibitemShut {NoStop}%
\bibitem [{\citenamefont {Bhatt}, \citenamefont {Zhang},\ and\ \citenamefont
  {Zuckerman}(2010)}]{Bhatt2010Steady-state-simulat}%
  \BibitemOpen
  \bibfield  {author} {\bibinfo {author} {\bibfnamefont {D.}~\bibnamefont
  {Bhatt}}, \bibinfo {author} {\bibfnamefont {B.~W.}\ \bibnamefont {Zhang}}, \
  and\ \bibinfo {author} {\bibfnamefont {D.~M.}\ \bibnamefont {Zuckerman}},\
  }\bibfield  {title} {\enquote {\bibinfo {title} {Steady-state simulations
  using weighted ensemble path sampling},}\ }\href {\doibase 10.1063/1.3456985}
  {\bibfield  {journal} {\bibinfo  {journal} {J Chem Phys}\ }\textbf {\bibinfo
  {volume} {133}},\ \bibinfo {pages} {014110} (\bibinfo {year}
  {2010})}\BibitemShut {NoStop}%
\bibitem [{\citenamefont {Rojnuckarin}, \citenamefont {Livesay},\ and\
  \citenamefont {Subramaniam}(2000)}]{Rojnuckarin2000Bimolecular-rea}%
  \BibitemOpen
  \bibfield  {author} {\bibinfo {author} {\bibfnamefont {A.}~\bibnamefont
  {Rojnuckarin}}, \bibinfo {author} {\bibfnamefont {D.~R.}\ \bibnamefont
  {Livesay}}, \ and\ \bibinfo {author} {\bibfnamefont {S.}~\bibnamefont
  {Subramaniam}},\ }\bibfield  {title} {\enquote {\bibinfo {title} {Bimolecular
  reaction simulation using {W}eighted {E}nsemble {B}rownian dynamics and the
  {U}niversity of {H}ouston {B}rownian {D}ynamics program},}\ }\href {\doibase
  10.1016/S0006-3495(00)76327-2} {\bibfield  {journal} {\bibinfo  {journal}
  {Biophys J}\ }\textbf {\bibinfo {volume} {79}},\ \bibinfo {pages} {686--93}
  (\bibinfo {year} {2000})}\BibitemShut {NoStop}%
\bibitem [{\citenamefont {Rojnuckarin}, \citenamefont {Kim},\ and\
  \citenamefont {Subramaniam}(1998)}]{Rojnuckarin1998Brownian-dynami}%
  \BibitemOpen
  \bibfield  {author} {\bibinfo {author} {\bibfnamefont {A.}~\bibnamefont
  {Rojnuckarin}}, \bibinfo {author} {\bibfnamefont {S.}~\bibnamefont {Kim}}, \
  and\ \bibinfo {author} {\bibfnamefont {S.}~\bibnamefont {Subramaniam}},\
  }\bibfield  {title} {\enquote {\bibinfo {title} {Brownian dynamics
  simulations of protein folding: access to milliseconds time scale and
  beyond},}\ }\href@noop {} {\bibfield  {journal} {\bibinfo  {journal} {Proc
  Natl Acad Sci U S A}\ }\textbf {\bibinfo {volume} {95}},\ \bibinfo {pages}
  {4288--92} (\bibinfo {year} {1998})}\BibitemShut {NoStop}%
\bibitem [{\citenamefont {Zwier}, \citenamefont {Kaus},\ and\ \citenamefont
  {Chong}(2011)}]{Zwier2011Efficient-Expli}%
  \BibitemOpen
  \bibfield  {author} {\bibinfo {author} {\bibfnamefont {M.}~\bibnamefont
  {Zwier}}, \bibinfo {author} {\bibfnamefont {J.}~\bibnamefont {Kaus}}, \ and\
  \bibinfo {author} {\bibfnamefont {L.}~\bibnamefont {Chong}},\ }\bibfield
  {title} {\enquote {\bibinfo {title} {Efficient explicit-solvent molecular
  dynamics simulations of molecular association kinetics: {M}ethane/methane,
  {N}a+/{C}l-, methane/benzene, and {K}+/18-crown-6 ether},}\ }\href@noop {}
  {\bibfield  {journal} {\bibinfo  {journal} {Journal of Chemical Theory and
  Computation}\ }\textbf {\bibinfo {volume} {7}},\ \bibinfo {pages}
  {1189--1197} (\bibinfo {year} {2011})}\BibitemShut {NoStop}%
\bibitem [{\citenamefont {Dickson}, \citenamefont {Warmflash},\ and\
  \citenamefont
  {Dinner}(2009{\natexlab{b}})}]{Dickson2009Separating-forward-a}%
  \BibitemOpen
  \bibfield  {author} {\bibinfo {author} {\bibfnamefont {A.}~\bibnamefont
  {Dickson}}, \bibinfo {author} {\bibfnamefont {A.}~\bibnamefont {Warmflash}},
  \ and\ \bibinfo {author} {\bibfnamefont {A.~R.}\ \bibnamefont {Dinner}},\
  }\bibfield  {title} {\enquote {\bibinfo {title} {Separating forward and
  backward pathways in nonequilibrium umbrella sampling},}\ }\href {\doibase
  10.1063/1.3244561} {\bibfield  {journal} {\bibinfo  {journal} {J Chem Phys}\
  }\textbf {\bibinfo {volume} {131}},\ \bibinfo {pages} {154104} (\bibinfo
  {year} {2009}{\natexlab{b}})}\BibitemShut {NoStop}%
\bibitem [{\citenamefont {Zhu}\ and\ \citenamefont
  {Hummer}(2010)}]{Zhu2010Pore-opening-and-clo}%
  \BibitemOpen
  \bibfield  {author} {\bibinfo {author} {\bibfnamefont {F.}~\bibnamefont
  {Zhu}}\ and\ \bibinfo {author} {\bibfnamefont {G.}~\bibnamefont {Hummer}},\
  }\bibfield  {title} {\enquote {\bibinfo {title} {Pore opening and closing of
  a pentameric ligand-gated ion channel},}\ }\href {\doibase
  10.1073/pnas.1009313107} {\bibfield  {journal} {\bibinfo  {journal} {Proc
  Natl Acad Sci U S A}\ }\textbf {\bibinfo {volume} {107}},\ \bibinfo {pages}
  {19814--9} (\bibinfo {year} {2010})}\BibitemShut {NoStop}%
\bibitem [{\citenamefont {Lettieri}\ \emph {et~al.}(2012)\citenamefont
  {Lettieri}, \citenamefont {Zwier}, \citenamefont {Stringer}, \citenamefont
  {Suarez}, \citenamefont {Chong},\ and\ \citenamefont
  {Zuckerman}}]{Lettieri2012Simultaneous-computa}%
  \BibitemOpen
  \bibfield  {author} {\bibinfo {author} {\bibfnamefont {S.}~\bibnamefont
  {Lettieri}}, \bibinfo {author} {\bibfnamefont {M.~C.}\ \bibnamefont {Zwier}},
  \bibinfo {author} {\bibfnamefont {C.~A.}\ \bibnamefont {Stringer}}, \bibinfo
  {author} {\bibfnamefont {E.}~\bibnamefont {Suarez}}, \bibinfo {author}
  {\bibfnamefont {L.~T.}\ \bibnamefont {Chong}}, \ and\ \bibinfo {author}
  {\bibfnamefont {D.~M.}\ \bibnamefont {Zuckerman}},\ }\href@noop {} {\enquote
  {\bibinfo {title} {Simultaneous computation of dynamical and equilibrium
  information using a weighted ensemble of trajectories},}\ } (\bibinfo {year}
  {2012}),\ \Eprint {http://arxiv.org/abs/1210.3094[physics.bio-ph]}
  {arXiv:1210.3094[physics.bio-ph]} \BibitemShut {NoStop}%
\bibitem [{\citenamefont {Vanden-Eijnden}\ and\ \citenamefont
  {Venturoli}(2009{\natexlab{b}})}]{Vanden-Eijnden2009Exact-rate-calculati}%
  \BibitemOpen
  \bibfield  {author} {\bibinfo {author} {\bibfnamefont {E.}~\bibnamefont
  {Vanden-Eijnden}}\ and\ \bibinfo {author} {\bibfnamefont {M.}~\bibnamefont
  {Venturoli}},\ }\bibfield  {title} {\enquote {\bibinfo {title} {Exact rate
  calculations by trajectory parallelization and tilting},}\ }\href {\doibase
  10.1063/1.3180821} {\bibfield  {journal} {\bibinfo  {journal} {J Chem Phys}\
  }\textbf {\bibinfo {volume} {131}},\ \bibinfo {pages} {044120} (\bibinfo
  {year} {2009}{\natexlab{b}})}\BibitemShut {NoStop}%
\bibitem [{\citenamefont {Dickson}\ \emph {et~al.}(2011)\citenamefont
  {Dickson}, \citenamefont {Maienschein-Cline}, \citenamefont {Tovo-Dwyer},
  \citenamefont {Hammond},\ and\ \citenamefont
  {Dinner}}]{Dickson2011Flow-dependent-}%
  \BibitemOpen
  \bibfield  {author} {\bibinfo {author} {\bibfnamefont {A.}~\bibnamefont
  {Dickson}}, \bibinfo {author} {\bibfnamefont {M.}~\bibnamefont
  {Maienschein-Cline}}, \bibinfo {author} {\bibfnamefont {A.}~\bibnamefont
  {Tovo-Dwyer}}, \bibinfo {author} {\bibfnamefont {J.}~\bibnamefont {Hammond}},
  \ and\ \bibinfo {author} {\bibfnamefont {A.}~\bibnamefont {Dinner}},\
  }\bibfield  {title} {\enquote {\bibinfo {title} {Flow-dependent unfolding and
  refolding of an {RNA} by nonequilibrium umbrella sampling},}\ }\href@noop {}
  {\bibfield  {journal} {\bibinfo  {journal} {Journal of Chemical Theory and
  Computation}\ }\textbf {\bibinfo {volume} {7}},\ \bibinfo {pages}
  {2710--2720} (\bibinfo {year} {2011})}\BibitemShut {NoStop}%
\bibitem [{wes()}]{west-pa}%
  \BibitemOpen
  \href@noop {} {}\bibinfo {howpublished}
  {\url{http://chong.chem.pitt.edu/WESTPA}}\BibitemShut {NoStop}%
\bibitem [{\citenamefont {Dickson}, \citenamefont {Warmflash},\ and\
  \citenamefont {Dinner}(2012{\natexlab{a}})}]{Dickson2012Erratum:-Nonequ}%
  \BibitemOpen
  \bibfield  {author} {\bibinfo {author} {\bibfnamefont {A.}~\bibnamefont
  {Dickson}}, \bibinfo {author} {\bibfnamefont {A.}~\bibnamefont {Warmflash}},
  \ and\ \bibinfo {author} {\bibfnamefont {A.~R.}\ \bibnamefont {Dinner}},\
  }\bibfield  {title} {\enquote {\bibinfo {title} {Erratum: "{N}onequilibrium
  umbrella sampling in spaces of many order parameters" [{J}. {C}hem. {P}hys.
  130, 074104 (2009)]},}\ }\href {\doibase 10.1063/1.4729744} {\bibfield
  {journal} {\bibinfo  {journal} {J Chem Phys}\ }\textbf {\bibinfo {volume}
  {136}},\ \bibinfo {pages} {229901} (\bibinfo {year}
  {2012}{\natexlab{a}})}\BibitemShut {NoStop}%
\bibitem [{\citenamefont {Dickson}, \citenamefont {Warmflash},\ and\
  \citenamefont {Dinner}(2012{\natexlab{b}})}]{Dickson2012Erratum:-Separa}%
  \BibitemOpen
  \bibfield  {author} {\bibinfo {author} {\bibfnamefont {A.}~\bibnamefont
  {Dickson}}, \bibinfo {author} {\bibfnamefont {A.}~\bibnamefont {Warmflash}},
  \ and\ \bibinfo {author} {\bibfnamefont {A.~R.}\ \bibnamefont {Dinner}},\
  }\bibfield  {title} {\enquote {\bibinfo {title} {Erratum: "{S}eparating
  forward and backward pathways in nonequilibrium umbrella sampling" [{J}.
  {C}hem. {P}hys. 131, 154104 (2009)]},}\ }\href {\doibase 10.1063/1.4730937}
  {\bibfield  {journal} {\bibinfo  {journal} {J Chem Phys}\ }\textbf {\bibinfo
  {volume} {136}},\ \bibinfo {pages} {239901} (\bibinfo {year}
  {2012}{\natexlab{b}})}\BibitemShut {NoStop}%
\bibitem [{\citenamefont {Damjanovi{\'c}}, \citenamefont
  {Garc{\'\i}a-Moreno~E},\ and\ \citenamefont
  {Brooks}(2009)}]{Damjanovic2009Self-guided-Langevin}%
  \BibitemOpen
  \bibfield  {author} {\bibinfo {author} {\bibfnamefont {A.}~\bibnamefont
  {Damjanovi{\'c}}}, \bibinfo {author} {\bibfnamefont {B.}~\bibnamefont
  {Garc{\'\i}a-Moreno~E}}, \ and\ \bibinfo {author} {\bibfnamefont {B.~R.}\
  \bibnamefont {Brooks}},\ }\bibfield  {title} {\enquote {\bibinfo {title}
  {Self-guided langevin dynamics study of regulatory interactions in
  {N}tr{C}},}\ }\href {\doibase 10.1002/prot.22439} {\bibfield  {journal}
  {\bibinfo  {journal} {Proteins}\ }\textbf {\bibinfo {volume} {76}},\ \bibinfo
  {pages} {1007--19} (\bibinfo {year} {2009})}\BibitemShut {NoStop}%
\bibitem [{\citenamefont {Lei}\ \emph {et~al.}(2009)\citenamefont {Lei},
  \citenamefont {Velos}, \citenamefont {Gardino}, \citenamefont {Kivenson},
  \citenamefont {Karplus},\ and\ \citenamefont
  {Kern}}]{Lei2009Segmented-transition}%
  \BibitemOpen
  \bibfield  {author} {\bibinfo {author} {\bibfnamefont {M.}~\bibnamefont
  {Lei}}, \bibinfo {author} {\bibfnamefont {J.}~\bibnamefont {Velos}}, \bibinfo
  {author} {\bibfnamefont {A.}~\bibnamefont {Gardino}}, \bibinfo {author}
  {\bibfnamefont {A.}~\bibnamefont {Kivenson}}, \bibinfo {author}
  {\bibfnamefont {M.}~\bibnamefont {Karplus}}, \ and\ \bibinfo {author}
  {\bibfnamefont {D.}~\bibnamefont {Kern}},\ }\bibfield  {title} {\enquote
  {\bibinfo {title} {Segmented transition pathway of the signaling protein
  nitrogen regulatory protein {C}},}\ }\href {\doibase
  10.1016/j.jmb.2009.06.065} {\bibfield  {journal} {\bibinfo  {journal} {J Mol
  Biol}\ }\textbf {\bibinfo {volume} {392}},\ \bibinfo {pages} {823--36}
  (\bibinfo {year} {2009})}\BibitemShut {NoStop}%
\bibitem [{\citenamefont {Adelman}\ \emph {et~al.}(2010)\citenamefont
  {Adelman}, \citenamefont {Chodera}, \citenamefont {Kuo}, \citenamefont
  {Miller},\ and\ \citenamefont {Barsky}}]{Adelman2010The-mechanical-prope}%
  \BibitemOpen
  \bibfield  {author} {\bibinfo {author} {\bibfnamefont {J.~L.}\ \bibnamefont
  {Adelman}}, \bibinfo {author} {\bibfnamefont {J.~D.}\ \bibnamefont
  {Chodera}}, \bibinfo {author} {\bibfnamefont {I.-F.~W.}\ \bibnamefont {Kuo}},
  \bibinfo {author} {\bibfnamefont {T.~F.}\ \bibnamefont {Miller},
  \bibfnamefont {3rd}}, \ and\ \bibinfo {author} {\bibfnamefont
  {D.}~\bibnamefont {Barsky}},\ }\bibfield  {title} {\enquote {\bibinfo {title}
  {The mechanical properties of {PCNA}: {I}mplications for the loading and
  function of a {DNA} sliding clamp},}\ }\href {\doibase
  10.1016/j.bpj.2010.03.056} {\bibfield  {journal} {\bibinfo  {journal}
  {Biophys J}\ }\textbf {\bibinfo {volume} {98}},\ \bibinfo {pages} {3062--9}
  (\bibinfo {year} {2010})}\BibitemShut {NoStop}%
\bibitem [{\citenamefont {Lyman}, \citenamefont {Pfaendtner},\ and\
  \citenamefont {Voth}(2008)}]{Lyman2008Systematic-multiscal}%
  \BibitemOpen
  \bibfield  {author} {\bibinfo {author} {\bibfnamefont {E.}~\bibnamefont
  {Lyman}}, \bibinfo {author} {\bibfnamefont {J.}~\bibnamefont {Pfaendtner}}, \
  and\ \bibinfo {author} {\bibfnamefont {G.~A.}\ \bibnamefont {Voth}},\
  }\bibfield  {title} {\enquote {\bibinfo {title} {Systematic multiscale
  parameterization of heterogeneous elastic network models of proteins},}\
  }\href {\doibase 10.1529/biophysj.108.139733} {\bibfield  {journal} {\bibinfo
   {journal} {Biophys J}\ }\textbf {\bibinfo {volume} {95}},\ \bibinfo {pages}
  {4183--92} (\bibinfo {year} {2008})}\BibitemShut {NoStop}%
\bibitem [{\citenamefont {Yang}\ \emph {et~al.}(2008)\citenamefont {Yang},
  \citenamefont {Song}, \citenamefont {Carriquiry},\ and\ \citenamefont
  {Jernigan}}]{Yang2008Close-correspondence}%
  \BibitemOpen
  \bibfield  {author} {\bibinfo {author} {\bibfnamefont {L.}~\bibnamefont
  {Yang}}, \bibinfo {author} {\bibfnamefont {G.}~\bibnamefont {Song}}, \bibinfo
  {author} {\bibfnamefont {A.}~\bibnamefont {Carriquiry}}, \ and\ \bibinfo
  {author} {\bibfnamefont {R.~L.}\ \bibnamefont {Jernigan}},\ }\bibfield
  {title} {\enquote {\bibinfo {title} {Close correspondence between the motions
  from principal component analysis of multiple {HIV}-1 protease structures and
  elastic network modes},}\ }\href {\doibase 10.1016/j.str.2007.12.011}
  {\bibfield  {journal} {\bibinfo  {journal} {Structure}\ }\textbf {\bibinfo
  {volume} {16}},\ \bibinfo {pages} {321--30} (\bibinfo {year}
  {2008})}\BibitemShut {NoStop}%
\bibitem [{\citenamefont {Kern}\ \emph {et~al.}(1999)\citenamefont {Kern},
  \citenamefont {Volkman}, \citenamefont {Luginb{\"u}hl}, \citenamefont
  {Nohaile}, \citenamefont {Kustu},\ and\ \citenamefont
  {Wemmer}}]{Kern1999Structure-of-a-trans}%
  \BibitemOpen
  \bibfield  {author} {\bibinfo {author} {\bibfnamefont {D.}~\bibnamefont
  {Kern}}, \bibinfo {author} {\bibfnamefont {B.~F.}\ \bibnamefont {Volkman}},
  \bibinfo {author} {\bibfnamefont {P.}~\bibnamefont {Luginb{\"u}hl}}, \bibinfo
  {author} {\bibfnamefont {M.~J.}\ \bibnamefont {Nohaile}}, \bibinfo {author}
  {\bibfnamefont {S.}~\bibnamefont {Kustu}}, \ and\ \bibinfo {author}
  {\bibfnamefont {D.~E.}\ \bibnamefont {Wemmer}},\ }\bibfield  {title}
  {\enquote {\bibinfo {title} {Structure of a transiently phosphorylated switch
  in bacterial signal transduction},}\ }\href {\doibase 10.1038/47273}
  {\bibfield  {journal} {\bibinfo  {journal} {Nature}\ }\textbf {\bibinfo
  {volume} {402}},\ \bibinfo {pages} {894--8} (\bibinfo {year}
  {1999})}\BibitemShut {NoStop}%
\bibitem [{\citenamefont {Theobald}(2005)}]{Theobald2005Rapid-calculation-of}%
  \BibitemOpen
  \bibfield  {author} {\bibinfo {author} {\bibfnamefont {D.~L.}\ \bibnamefont
  {Theobald}},\ }\bibfield  {title} {\enquote {\bibinfo {title} {Rapid
  calculation of {RMSD}s using a quaternion-based characteristic polynomial},}\
  }\href {\doibase 10.1107/S0108767305015266} {\bibfield  {journal} {\bibinfo
  {journal} {Acta Crystallogr A}\ }\textbf {\bibinfo {volume} {61}},\ \bibinfo
  {pages} {478--80} (\bibinfo {year} {2005})}\BibitemShut {NoStop}%
\bibitem [{\citenamefont {Chodera}\ \emph {et~al.}(2007)\citenamefont
  {Chodera}, \citenamefont {Swope}, \citenamefont {Pitera}, \citenamefont
  {Seok},\ and\ \citenamefont {Dill}}]{Chodera2007Use-of-the-weighted-}%
  \BibitemOpen
  \bibfield  {author} {\bibinfo {author} {\bibfnamefont {J.}~\bibnamefont
  {Chodera}}, \bibinfo {author} {\bibfnamefont {W.}~\bibnamefont {Swope}},
  \bibinfo {author} {\bibfnamefont {J.}~\bibnamefont {Pitera}}, \bibinfo
  {author} {\bibfnamefont {C.}~\bibnamefont {Seok}}, \ and\ \bibinfo {author}
  {\bibfnamefont {K.}~\bibnamefont {Dill}},\ }\bibfield  {title} {\enquote
  {\bibinfo {title} {Use of the weighted histogram analysis method for the
  analysis of simulated and parallel tempering simulations},}\ }\href@noop {}
  {\bibfield  {journal} {\bibinfo  {journal} {Journal of Chemical Theory and
  Computation}\ }\textbf {\bibinfo {volume} {3}},\ \bibinfo {pages} {26--41}
  (\bibinfo {year} {2007})}\BibitemShut {NoStop}%
\bibitem [{\citenamefont {Shirts}\ and\ \citenamefont
  {Chodera}(2008)}]{Shirts2008Statistically-optima}%
  \BibitemOpen
  \bibfield  {author} {\bibinfo {author} {\bibfnamefont {M.~R.}\ \bibnamefont
  {Shirts}}\ and\ \bibinfo {author} {\bibfnamefont {J.~D.}\ \bibnamefont
  {Chodera}},\ }\bibfield  {title} {\enquote {\bibinfo {title} {Statistically
  optimal analysis of samples from multiple equilibrium states},}\ }\href
  {\doibase 10.1063/1.2978177} {\bibfield  {journal} {\bibinfo  {journal} {J
  Chem Phys}\ }\textbf {\bibinfo {volume} {129}},\ \bibinfo {pages} {124105}
  (\bibinfo {year} {2008})}\BibitemShut {NoStop}%
\bibitem [{\citenamefont {Du}\ \emph {et~al.}(1998)\citenamefont {Du},
  \citenamefont {Pande}, \citenamefont {Grosberg}, \citenamefont {Tanaka},\
  and\ \citenamefont {Shakhnovich}}]{Du1998On-the-transiti}%
  \BibitemOpen
  \bibfield  {author} {\bibinfo {author} {\bibfnamefont {R.}~\bibnamefont
  {Du}}, \bibinfo {author} {\bibfnamefont {V.}~\bibnamefont {Pande}}, \bibinfo
  {author} {\bibfnamefont {A.}~\bibnamefont {Grosberg}}, \bibinfo {author}
  {\bibfnamefont {T.}~\bibnamefont {Tanaka}}, \ and\ \bibinfo {author}
  {\bibfnamefont {E.}~\bibnamefont {Shakhnovich}},\ }\bibfield  {title}
  {\enquote {\bibinfo {title} {On the transition coordinate for protein
  folding},}\ }\href@noop {} {\bibfield  {journal} {\bibinfo  {journal} {The
  Journal of chemical physics}\ }\textbf {\bibinfo {volume} {108}},\ \bibinfo
  {pages} {334} (\bibinfo {year} {1998})}\BibitemShut {NoStop}%
\bibitem [{\citenamefont {Geissler}, \citenamefont {Dellago},\ and\
  \citenamefont {Chandler}(1999)}]{Geissler1999Kinetic-pathways-of-}%
  \BibitemOpen
  \bibfield  {author} {\bibinfo {author} {\bibfnamefont {P.}~\bibnamefont
  {Geissler}}, \bibinfo {author} {\bibfnamefont {C.}~\bibnamefont {Dellago}}, \
  and\ \bibinfo {author} {\bibfnamefont {D.}~\bibnamefont {Chandler}},\
  }\bibfield  {title} {\enquote {\bibinfo {title} {Kinetic pathways of ion pair
  dissociation in water},}\ }\href@noop {} {\bibfield  {journal} {\bibinfo
  {journal} {The Journal of Physical Chemistry B}\ }\textbf {\bibinfo {volume}
  {103}},\ \bibinfo {pages} {3706--3710} (\bibinfo {year} {1999})}\BibitemShut
  {NoStop}%
\bibitem [{\citenamefont {No{\'e}}\ \emph {et~al.}(2009)\citenamefont
  {No{\'e}}, \citenamefont {Sch{\"u}tte}, \citenamefont {Vanden-Eijnden},
  \citenamefont {Reich},\ and\ \citenamefont {Weikl}}]{Noe2009Constructing-th}%
  \BibitemOpen
  \bibfield  {author} {\bibinfo {author} {\bibfnamefont {F.}~\bibnamefont
  {No{\'e}}}, \bibinfo {author} {\bibfnamefont {C.}~\bibnamefont
  {Sch{\"u}tte}}, \bibinfo {author} {\bibfnamefont {E.}~\bibnamefont
  {Vanden-Eijnden}}, \bibinfo {author} {\bibfnamefont {L.}~\bibnamefont
  {Reich}}, \ and\ \bibinfo {author} {\bibfnamefont {T.~R.}\ \bibnamefont
  {Weikl}},\ }\bibfield  {title} {\enquote {\bibinfo {title} {Constructing the
  equilibrium ensemble of folding pathways from short off-equilibrium
  simulations},}\ }\href {\doibase 10.1073/pnas.0905466106} {\bibfield
  {journal} {\bibinfo  {journal} {Proc Natl Acad Sci U S A}\ }\textbf {\bibinfo
  {volume} {106}},\ \bibinfo {pages} {19011--6} (\bibinfo {year}
  {2009})}\BibitemShut {NoStop}%
\bibitem [{\citenamefont {Prinz}\ \emph {et~al.}(2011)\citenamefont {Prinz},
  \citenamefont {Wu}, \citenamefont {Sarich}, \citenamefont {Keller},
  \citenamefont {Senne}, \citenamefont {Held}, \citenamefont {Chodera},
  \citenamefont {Sch{\"u}tte},\ and\ \citenamefont
  {No{\'e}}}]{Prinz2011Markov-models-o}%
  \BibitemOpen
  \bibfield  {author} {\bibinfo {author} {\bibfnamefont {J.-H.}\ \bibnamefont
  {Prinz}}, \bibinfo {author} {\bibfnamefont {H.}~\bibnamefont {Wu}}, \bibinfo
  {author} {\bibfnamefont {M.}~\bibnamefont {Sarich}}, \bibinfo {author}
  {\bibfnamefont {B.}~\bibnamefont {Keller}}, \bibinfo {author} {\bibfnamefont
  {M.}~\bibnamefont {Senne}}, \bibinfo {author} {\bibfnamefont
  {M.}~\bibnamefont {Held}}, \bibinfo {author} {\bibfnamefont {J.~D.}\
  \bibnamefont {Chodera}}, \bibinfo {author} {\bibfnamefont {C.}~\bibnamefont
  {Sch{\"u}tte}}, \ and\ \bibinfo {author} {\bibfnamefont {F.}~\bibnamefont
  {No{\'e}}},\ }\bibfield  {title} {\enquote {\bibinfo {title} {Markov models
  of molecular kinetics: generation and validation},}\ }\href {\doibase
  10.1063/1.3565032} {\bibfield  {journal} {\bibinfo  {journal} {J Chem Phys}\
  }\textbf {\bibinfo {volume} {134}},\ \bibinfo {pages} {174105} (\bibinfo
  {year} {2011})}\BibitemShut {NoStop}%
\bibitem [{\citenamefont {Maragliano}, \citenamefont {Vanden-Eijnden},\ and\
  \citenamefont {Roux}(2009)}]{Maragliano2009Free-energy-and}%
  \BibitemOpen
  \bibfield  {author} {\bibinfo {author} {\bibfnamefont {L.}~\bibnamefont
  {Maragliano}}, \bibinfo {author} {\bibfnamefont {E.}~\bibnamefont
  {Vanden-Eijnden}}, \ and\ \bibinfo {author} {\bibfnamefont {B.}~\bibnamefont
  {Roux}},\ }\bibfield  {title} {\enquote {\bibinfo {title} {Free energy and
  kinetics of conformational transitions from voronoi tessellated milestoning
  with restraining potentials},}\ }\href {\doibase 10.1021/ct900279z}
  {\bibfield  {journal} {\bibinfo  {journal} {J Chem Theory Comput}\ }\textbf
  {\bibinfo {volume} {5}},\ \bibinfo {pages} {2589--2594} (\bibinfo {year}
  {2009})}\BibitemShut {NoStop}%
\bibitem [{\citenamefont {Peters}\ \emph {et~al.}(2004)\citenamefont {Peters},
  \citenamefont {Heyden}, \citenamefont {Bell},\ and\ \citenamefont
  {Chakraborty}}]{Peters2004A-growing-string-met}%
  \BibitemOpen
  \bibfield  {author} {\bibinfo {author} {\bibfnamefont {B.}~\bibnamefont
  {Peters}}, \bibinfo {author} {\bibfnamefont {A.}~\bibnamefont {Heyden}},
  \bibinfo {author} {\bibfnamefont {A.~T.}\ \bibnamefont {Bell}}, \ and\
  \bibinfo {author} {\bibfnamefont {A.}~\bibnamefont {Chakraborty}},\
  }\bibfield  {title} {\enquote {\bibinfo {title} {A growing string method for
  determining transition states: comparison to the nudged elastic band and
  string methods},}\ }\href {\doibase 10.1063/1.1691018} {\bibfield  {journal}
  {\bibinfo  {journal} {J Chem Phys}\ }\textbf {\bibinfo {volume} {120}},\
  \bibinfo {pages} {7877--86} (\bibinfo {year} {2004})}\BibitemShut {NoStop}%
\end{thebibliography}

%

\clearpage

\clearpage

%
\begin{table}
\caption{
\label{tab:2dper_params}
Parameters used in the WE simulations for the periodic two-dimensional potential.}
\begin{ruledtabular}
\begin{tabular}{lcc}
 & Common & Rare  \\
\cline{2-3}
$\alpha$ & 1.125 & 2.25 \\
$N_{\text{im}}$  & 20  & 50 \\
$N_{\text{rep}}$ & 40  & 50   \\
$\tau/{\delta}t$ & 10 & 10 \\
\\
\cline{1-1}
Iterations ($\tau$) per phase \\
\cline{1-1}
Phase I & 1000 & 5000 \\
Phase II & 34000 & 30000 \\
\end{tabular}
\begin{tabular}{lcc}
& Phase I & Phase II \\
$T_{\text{move}}/\tau$ & 25 & -- \\
$T_{\text{avg}}/\tau$ & 100 & -- \\
$P$ & 2 & --  \\
\end{tabular}
\end{ruledtabular}
\end{table}
%

\begin{table}
\caption{
\label{tab:2drp_params}
Parameters used in the WE simulations for the two-dimensional ring potential.}
\begin{ruledtabular}
\begin{tabular}{lccccc}
$\beta$ & 1.0 & 1.5 & 2.0 & 2.5 & 3.0 \\
\cline{2-6}
$N_{\text{im}}$  & 50  & 60 & 80 & 100  & 100\\
$N_{\text{rep}}$ & 40  & 50  & 50 & 50  &   50 \\
$\tau/{\delta}t$ & 10 & 10 & 10 & 10 & 10 \\
\\
\cline{1-1}
Iterations ($\tau$) per phase \\
\cline{1-1}
Phase I & 800 & 800 & 800 & 800 & 800\\
Phase II & 700 & 1700 & 1700 & 3200 & 4200 \\
Phase III & 3500 & 7500 & 7500 & 6000 & 20000 \\
\end{tabular}
\begin{tabular}{lccc}
& Phase I & Phase II & Phase III \\
$T_{\text{move}}/\tau$ & 10 & -- & -- \\
$T_{\text{avg}}/\tau$ & 50 & -- & -- \\
$P$ & 2 & -- & --  \\
$T_{\text{rw}}/\tau$ & 20 & 20 & -- \\
$T_{\text{ravg}}/N_{\tau}\tau$ & 0.5 & 0.5 & -- \\

\end{tabular}
\end{ruledtabular}
\end{table}

\end{document}